\pdfoutput=1
\documentclass[sigconf, nonacm, balance=false]{acmart}

\AtBeginDocument{%
  }

\usepackage{amsmath}
\usepackage{multirow}
\usepackage{enumitem}
\usepackage{algorithm}
\usepackage{algpseudocode}

\begin{document}

\title{Beyond Long-Tail POIs: Transition-Centered Modeling for Human
  Mobility Prediction}

\author{Dingyang Lyu}
\affiliation{\institution{The University of Melbourne}\city{Melbourne}\country{Australia}}
\email{dingyang.lyu@student.unimelb.edu.au}

\author{Zhengjia Xu}
\affiliation{\institution{Macquarie University}\city{Sydney}\country{Australia}}
\email{zhengjia.xu@hdr.mq.edu.au}

\author{Jey Han Lau}
\affiliation{\institution{The University of Melbourne}\city{Melbourne}\country{Australia}}
\email{jeyhan.lau@unimelb.edu.au}

\author{Jianzhong Qi}
\affiliation{\institution{The University of Melbourne}\city{Melbourne}\country{Australia}}
\email{jianzhong.qi@unimelb.edu.au}

\begin{abstract}
Human mobility prediction, a core task of location-based web services, forecasts a user's next Point of Interest~(POI) visit from historical trajectories, supporting applications from recommendation to urban planning.
A general perception in the literature has been that visits to popular POIs are relatively easy to predict. 
Our analysis challenges this perception and shows that, even for visits to popular POIs, many predictions fail. The underlying cause is often transition-level sparsity, where a transition is a move from a source POI to a destination POI. While a POI may be popular, its transitions from other POIs may still be very sparse, since there are many POIs in a dataset. As a result, predicting a popular next POI can still be challenging if the transition from the source POI to it never appears in the training data.
We therefore argue that a core bottleneck in human mobility prediction lies in the transition-level long tail, and propose \textbf{RE}trieval \textbf{C}ombined with p\textbf{A}rametric memory for next-\textbf{P}OI prediction~(\textbf{RECAP}). RECAP learns frequent transitions in the parameters of a small Transformer trained from scratch. Long-tail transitions give the parameters little or no training signal.  RECAP retrieves evidence for them from two sources: (i) multi-hop transitivity learned from a global transition graph of all users, and (ii) revisit evidence from a target user's historical trajectory. RECAP further uses two-stage training to let the revisit component also learn from frequently seen transitions. Experiments on three real-world check-in datasets show that RECAP improves prediction accuracy, with clear gains on tail transitions. 
Our code is available at \url{https://github.com/NiallLDY/RECAP}.
\end{abstract}


\maketitle

\section{Introduction}
\label{sec:introduction}

Location-based services, such as check-in services, map search, and local business recommendation, are among the fundamental applications on the Web. Human mobility prediction is an important task behind location-based  services~\cite{deepmove,flashback,lotnext}. It predicts the next Point of Interest (POI) a user will visit from the user's check-in history, supporting POI recommendation, trip planning, and urban planning~\cite{stan,mclp}.
Progress in this sequential prediction problem has been driven by developments in sequence encoders, e.g., from Recurrent Neural Networks (RNNs)~\cite{strnn,deepmove} to Attention models~\cite{stan,mtnet,rotan}. 
Recent studies highlight the problem caused by long-tail POIs~\cite{lotnext,aloha,depoi}, where a small set of POIs receives most visits while a large fraction of POIs have reported few visits. These studies have proposed training strategies that reduce the dominance of popular POIs and let rarely visited POIs borrow signals from related POIs.

\begin{figure}[t]
    \centering
    \includegraphics[width=0.90\columnwidth]{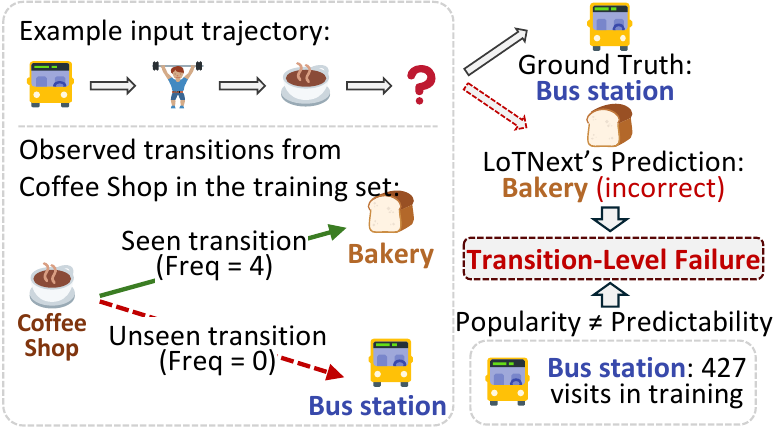}
    \caption{An example of prediction failure of LoTNext on a popular destination due to an unseen transition on NYC.}
    \label{fig:lotnext_example}
\end{figure}

We observe that prediction failures can occur even for highly popular POIs. As shown in Figure~\ref{fig:lotnext_example}, LoTNext~\cite{lotnext}, a state-of-the-art (SOTA) long-tail model, fails to predict \texttt{Bus Station} despite its high visit frequency in the NYC~\cite{nyctky} training set, ranking at the 99th percentile.
In this case, the ground-truth transition \texttt{Coffee Shop} $\rightarrow$ \texttt{Bus Station} is absent from the training set, while the incorrectly predicted transition \texttt{Coffee Shop} $\rightarrow$ \texttt{Bakery} appears multiple times. This example exposes a \emph{transition-level} failure mode: the destination POI is frequent, yet its pairing with the current source POI lacks direct training support.

\begin{figure}[t]
    \centering
    \includegraphics[width=\columnwidth,trim={0 0 0 0.5bp},clip]{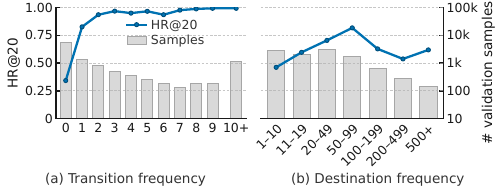}
    \caption{Accuracy aligns more with transition frequency than destination frequency.}
    \label{fig:freq_analysis}
\end{figure}

Figure~\ref{fig:freq_analysis} further shows that prediction accuracy correlates with transition frequency better than destination POI frequency.
It groups NYC validation samples by how often their transition (a) or destination POI (b) appears in training, and shows the sample count (bars) and the HR@20 of LoTNext (line) for each group.
Transitions that appear frequently in training have substantially higher accuracy, whereas low-frequency or unseen transitions are much more difficult to predict.
The counts of pairs in different transition frequency bins (bars in Figure~\ref{fig:freq_analysis}a) also reveal a highly long-tailed transition distribution: a large fraction of validation transitions appear only a few times in training, or never appear at all (i.e., transition frequency $= 0$).
These findings motivate a shift of the long-tail formulation from the item (i.e., POI) level to the transition level, making the prediction of long-tail transitions a central task.

\textbf{Challenges.} Model parameters can learn from the training occurrences of frequent transitions, and what they learn forms a parametric memory~\cite{feldman2020,knnlm}, but tail transitions give the parameters little or no training signal. Although a tail transition itself is rarely observed, its destination often appears in the user's earlier check-ins~\cite{song2010} (we observe this for 72.3\% of tail transitions on the NYC validation set), and other users' trajectories often link its source and destination through other POIs. The model therefore needs to be guided to look up and learn from this evidence. We thus view the transition-level long tail as a problem of combining parametric memory with retrieval, which raises two coupled challenges.
\textbf{(1)} \textbf{How to retrieve evidence for tail transitions:} The target POI of a tail transition may have appeared in the user's history, and the user now revisits it from a different source POI. We refer to these cases as \emph{user history revisits}. The target may also be connected to the current source POI by other users' transitions through multi-hop paths, and we refer to these cases as \emph{transitive completions}. Many existing methods already use the user's history or collaborative signals, for example through attention over past check-ins~\cite{deepmove,flashback} or global transition graphs~\cite{getnext,sthgcn}, yet their accuracy on tail transitions remains low. How to effectively retrieve and use this evidence is therefore a challenge.
\textbf{(2)} \textbf{How to balance parametric memory and retrieval:} Retrieved evidence should not override the parametric memory unconditionally, because whether a user returns to a visited POI depends on the current context. For example, the office may be the most frequent destination in a user's history and a popular transition from the current POI, yet the user is unlikely to return there an hour after leaving it. The revisit evidence must therefore be judged under the current context before it adjusts the prediction. Furthermore, training such a component is difficult. Frequent transitions dominate the training data and the parametric memory already predicts most of them, so a revisit component trained together with it through the final prediction can receive little training signal.

\textbf{Our Contributions.} We address these challenges and propose \textbf{RE}trieval \textbf{C}ombined with p\textbf{A}rametric memory for next-\textbf{P}OI prediction~(\textbf{RECAP}) to handle the transition-level long tail. RECAP learns frequent transitions in the parameters of a small Transformer trained from scratch, and retrieves transition evidence for the others from two sources. For user history revisits, the main source of its gains, it exposes revisit candidates from the user's historical trajectory and scores them under the current context. For destinations outside the user's history, it uses multi-hop paths in the transition graph to recall plausible POIs reached from the current source POI through intermediate POIs. We further use two-stage training, so that frequent transitions in the user's history also train the revisit scorer. To our knowledge, this is the first work in next POI prediction that defines the long-tail problem by transition frequency and models it explicitly. Our main contributions are:

\begin{itemize}[itemsep=2pt,leftmargin=12pt,topsep=0pt,parsep=0pt]
\item We formulate next POI prediction from a transition-level perspective, identify the transition long-tail problem, and evaluate prediction methods on transitions grouped by how often they appear in training.

\item We introduce RECAP, which combines parametric memory with retrieved user-history revisit signals and multi-hop transition-graph signals, and uses two-stage training so that the revisit scorer also learns from frequent transitions.

\item On real-world check-in datasets, the lightweight RECAP outperforms SOTA methods in overall hit rate and ranking quality, with an average improvement of 20.1\% over the best baselines on tail transitions, and is competitive with LLM-based methods while requiring far less inference time and GPU memory.
\end{itemize}

\section{Related Work}
\label{sec:related}

\paragraph{Next POI Prediction.}
Existing methods can be grouped into sequential and graph-based.
Sequential methods treat each user's check-ins as an ordered trajectory and learn POI visit (i.e., check-in) patterns with RNN or attention-based models.
DeepMove~\cite{deepmove} adds historical attention to recurrent trajectory encoding, RepeatNet~\cite{repeatnet} chooses between items in the current session and new items with a repeat-explore mechanism, Flashback~\cite{flashback} and REPLAY~\cite{replay} retrieve relevant past hidden states under similar spatio-temporal contexts, and DPRL~\cite{dprl} separates POI and region sequences into recurrent branches.
Attention models replace recurrent propagation with attention over historical check-ins, using multi-level temporal/context attention~\cite{multi_level_attention}, geography-aware self-attention~\cite{geography_aware}, bi-layer spatio-temporal attention~\cite{stan}, or rotation-based temporal attention for time-specific recommendation~\cite{rotan}. CaST-POI~\cite{castpoi} uses each candidate POI as the query of attention over the trajectory and adds an explicit revisit gate.
Graph-based methods share mobility signals across users and locations. They encode global trajectory-flow graphs~\cite{getnext}, trajectory-level hypergraphs~\cite{sthgcn}, multi-view hypergraphs over collaborative, transitional, and geographical relations~\cite{dchl}, or scenario-aware hypergraphs~\cite{msahg}.

With the recent rise of Large Language Models (LLMs), some generative methods have emerged. Early studies~\cite{llm-mob,llm4poi,nextlocllm} write trajectories as a text prompt and ask an LLM to generate the ID of the next POI. Later studies~\cite{gnpr_sid,geogr,r3vae} replace random POI IDs with \emph{semantic IDs}, and recent studies~\cite{agentmove,comapoi,mas4poi} build agent frameworks with external tools and knowledge.
Despite progress in mobility modeling, most existing methods focus on general next-POI accuracy, with limited attention to long-tailed mobility patterns.

Human mobility studies show that people mostly return to places they have visited: the probability of visiting a place grows with its past visits~\cite{song2010}, only about 20--25\% of moves go to new places~\cite{cuttone2018}, and a person keeps about 25 familiar places at any time~\cite{alessandretti2018}. Check-in datasets such as Foursquare and Gowalla also lack a sequential structure~\cite{k1poi}, as shuffling a user's check-ins barely changes the accuracy of sequential recommenders~\cite{klenitskiy2024}. These findings motivate RECAP to explicitly use statistics of the user's history to help predict tail transitions, rather than only feeding it to a sequence encoder.

\paragraph{Long-tail Learning.}
The long-tail problem has been widely studied in recommender systems, where tail items receive few user interactions and are difficult to learn. Existing solutions improve tail-item recommendation through popularity debiasing, graph augmentation, or adaptive training objectives~\cite{bias_debias_survey,pda,galore,melt}. 

In human mobility prediction, recent studies address this problem mainly at the destination-POI level: LoTNext~\cite{lotnext} adjusts graph learning and loss functions for long-tailed POIs. DePOI~\cite{depoi} separates causal mobility signals from biased graph signals. ALOHA~\cite{aloha} builds adaptive location hierarchies for long-tail POIs, and CTMR~\cite{ctmr} uses multimodal cross-task region information for long-tail POIs. These methods improve tail or debiased POI prediction, while their long-tail formulation is still defined mainly by destination POI visit frequency. SGRec~\cite{sgrec} notes that sparse POI-to-POI transitions limit sequential models and supplements them with the sequences of other users, but it does not define the long tail by transition frequency or evaluate methods on rare and unseen transitions. Our study fills the gap in studying long-tail transitions.

Closest to our perspective, a recent generative recommendation study~\cite{meta} analyzes memorization and generalization through unseen item transitions and uses a simple adaptive ensemble of item-ID and semantic-ID models to cover more transition combinations. It focuses on diagnosing this trade-off in general recommendation. We instead study transition sparsity in mobility prediction and improve tail transitions with transferable spatio-temporal signals.

\section{Method}
\label{sec:method}

Let $\mathcal U$ and $\mathcal P$ denote the sets of users and POIs, respectively.
A check-in $(u,p,\tau)$ denotes that user $u\in\mathcal U$ visits POI $p\in\mathcal P$ at timestamp $\tau$.
Each POI $p$ is associated with known metadata, including its ID, category $\operatorname{cat}(p)$, and geographical coordinates $\operatorname{loc}(p)$.
For each user $u$, we sort all check-ins by time and obtain a complete check-in sequence of length $L_u$, $Q_u=\langle (p^u_1,\tau^u_1),\ldots,(p^u_{L_u},\tau^u_{L_u})\rangle$, where $\tau^u_1<\cdots<\tau^u_{L_u}$.
Following prior work~\cite{getnext,im-poi}, we further split $Q_u$ by time gaps into a set of $J_u$ trajectories
$\mathcal S_u=\{S^u_1,\ldots,S^u_{J_u}\}$, where each $S^u_j$ is a subsequence of $Q_u$.
At prediction step $t<L_u$, the model observes the user's \emph{observed history}
$\mathcal H^u_t= \langle (p^u_1,\tau^u_1),\ldots,(p^u_t,\tau^u_t) \rangle$,
which contains all check-ins of user $u$ before the target check-in.
The prediction target is the next POI $y^u_t=p^u_{t+1}$, and the current source POI is $s^u_t=p^u_t$.
Each prediction therefore corresponds to a source--destination transition $(s^u_t,y^u_t)$.
When the context is clear, we omit the superscript $u$ and write $\mathcal H_t$, $p_t$, $\tau_t$, $s_t$, and $y_t$.

\paragraph{Transition-Level Long Tail.}
For a source-destination pair $(s,d)$, let $m_{sd}$ denote its occurrence count as adjacent check-ins in the training data.
For a test transition $(s_t,y_t)$, we call it \emph{unseen} when $m_{s_t y_t}=0$.
Given a threshold $\eta$, we call it a \emph{tail transition} when $m_{s_t y_t}\leq \eta$.
In this work, we set $\eta=1$, so tail transitions appear at most once in the training data, and \emph{head transitions} appear at least twice. We analyze the choice of $\eta$ in Appendix~\ref{app:tail-definition-sensitivity}.

\begin{figure*}[t]
    \centering
    \includegraphics[width=0.95\textwidth]{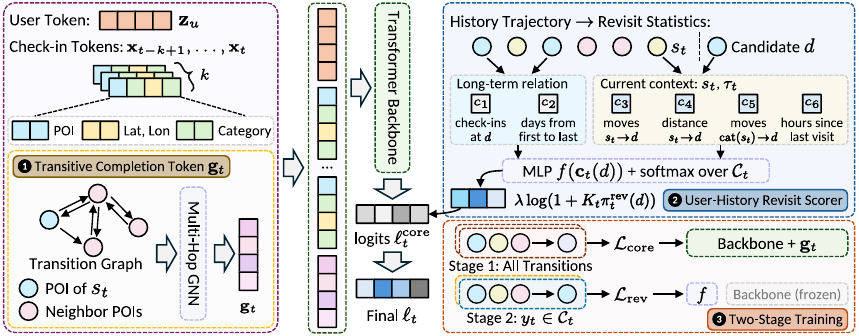}
    \caption{RECAP combines \ding{182} transitive completion token $\mathbf g_t$, built by multi-hop propagation on the transition graph, as part of the input, \ding{183} a user-history revisit scorer that adjusts the logits of the POIs the user has visited by scoring six statistics with a shared MLP $f$, and \ding{184} two-stage training to optimize the full framework, whose second stage trains only $f$ on cases with $y_t\in\mathcal C_t$.}
    \label{fig:recap}
\end{figure*}

\subsection{Overall Pipeline}
To address the long-tail transition issue, RECAP learns frequent transitions with a Transformer backbone and adds three main components as shown in Figure~\ref{fig:recap}: (1) multi-hop transition-graph encoding for finding plausible next POIs through observed POI-to-POI \emph{paths} in the global transition graph, (2) a user-history revisit scorer that adjusts the scores of the POIs the user has visited under the current context, and (3) two-stage training, which first trains the backbone and then trains the revisit scorer with the backbone frozen. The first two components provide the two sources of retrieved evidence for source--destination pairs rare in training: the revisit scorer reads the user's transitions from the current source to each visited POI, and the graph token reads multi-hop paths from the source. 

\paragraph{Trajectory Tokenization.}
RECAP tokenizes the input trajectory and encodes it with a Transformer encoder to produce the prediction logits. Given the observed history $\mathcal H_t$, we use its most recent length-$k$ (hyperparameter, Appendix~\ref{app:input-window-length}) suffix $(p_{t-k+1},\dots,p_t)$ as the backbone sequence, where $p_t$ is the current source POI $s_t$.
Each check-in is represented by a token that combines learnable embeddings of its POI ID and category ID with its coordinates:
\begin{equation}
\label{eq:checkin-token}
\mathbf x_q = \operatorname{MLP}_{\mathrm{tok}}
\big(
[
\mathbf e_{p_q};
\mathbf a_{p_q};
\operatorname{loc}(p_q)
]
\big),\quad q=t-k+1,\dots,t,
\end{equation}
where $\mathbf e_{p_q}$ is the learnable POI-ID embedding of POI $p_q$, $\mathbf a_{p_q}$ is the learnable category-ID embedding, and $\operatorname{MLP}_{\mathrm{tok}}$ maps the concatenated features to the Transformer hidden dimension.

\paragraph{Transformer Encoding and POI Scoring.}
We feed the token sequence into a Transformer encoder to produce the prediction state. The input token sequence includes the multi-hop transition signal (i.e., graph token $\mathbf g_t$) from \ding{182} in Figure~\ref{fig:recap}, which is detailed in Section~\ref{sec:multi_hop_transition_graph_encoding}.
We use the learnable user-ID embedding $\mathbf z_u$ as the user token for user $u$.
The Transformer input concatenates this user token, the check-in tokens (from Equation~\ref{eq:checkin-token}), and the graph token to produce the prediction state $\mathbf h_t$:
\begin{align}
\mathbf Z_t &=
[\mathbf z_u;\ \mathbf x_{t-k+1};\ \dots;\ \mathbf x_t;\ \mathbf g_t] + \mathbf E_{\mathrm{pos}},
\label{eq:encoder-input} \\
\mathbf h_t &=
\mathrm{TransformerEncoder}(\mathbf Z_t).
\label{eq:summary-state}
\end{align}
where $\mathbf E_{\mathrm{pos}}$ is a learnable positional embedding.
We then score each candidate POI $d\in\mathcal P$ by
\begin{equation}
\label{eq:core-score}
\ell_t^{\mathrm{core}}(d)=\mathbf w_d^\top \mathbf h_t+b_d,\qquad d\in\mathcal P,
\end{equation}
where $\mathbf w_d$ and $b_d$ are the output weight vector and bias for candidate POI $d$. The resulting score ranks all candidate POIs using the current trajectory state and the multi-hop transition signal.
Frequent transitions in training form the parametric memory of RECAP: the training cases of a transition from $s$ to $d$ raise the score of $d$ when $s$ is the current POI. Tail transitions have few such cases and remain hard to predict.

After obtaining the core logits, RECAP further scores the user's historical visited POIs and adjusts their logits (cf.~Section~\ref{sec:user_history_revisit_scorer} and \ding{183} in Figure~\ref{fig:recap}). Finally, RECAP trains the revisit scorer after the backbone, with the backbone frozen (cf.~Section~\ref{sec:two_stage_training} and \ding{184} in Figure~\ref{fig:recap}).

\subsection{Multi-Hop Transition-Graph Encoding}
\label{sec:multi_hop_transition_graph_encoding}
Tail transitions often lack direct source--destination evidence.
For destinations outside the observed history, RECAP retrieves evidence from other users: \emph{transitive completions}, that is, multi-hop paths in the global transition graph that start from the current source POI. We encode these paths into a source-specific graph token $\mathbf g_t$ for the Transformer backbone.

\paragraph{Transitive Completion Token.}
A transition graph~\cite{getnext} is built from the training data.
Each node is a POI, and each directed edge records how often one POI is followed by another in observed trajectories.
We define the row-normalized training-transition adjacency as:
\begin{equation}
\label{eq:transition-adj}
\mathbf A_{sd}=
\frac{m_{sd}}
{\sum_{d'\in\mathcal P} m_{sd'}},
\end{equation}
where $m_{sd}$ is the occurrence count of $(s, d)$ transitions in the training set as defined earlier.
Let $\mathbf E_{\mathrm{poi}}\in\mathbb R^{|\mathcal P|\times d_p}$ denote the matrix formed by stacking all learnable POI-ID embeddings $\{\mathbf e_p\}_{p\in\mathcal P}$ of dimension $d_p$, and initialize graph propagation as $\mathbf G^{(0)}=\mathbf E_{\mathrm{poi}}$.
We perform $N$-hop (hyperparameter) propagation on the graph:
\begin{equation}
\label{eq:graph-propagation}
\mathbf G^{(i+1)} = \mathbf A \mathbf G^{(i)}, \qquad i=0,\dots,N-1.
\end{equation}
Equation~\eqref{eq:graph-propagation} aggregates structural evidence from increasingly distant neighborhoods via multi-hop propagation.
For the current source $s_t$, we map its propagated representation to a graph token:
\begin{equation}
\label{eq:graph-token}
\mathbf g_t = \mathrm{LN}\!\big(\operatorname{MLP}_{\mathrm{graph}}(\mathbf G^{(N)}_{s_t})\big),
\end{equation}
where $\mathbf G^{(N)}_{s_t}$ is the row of $\mathbf G^{(N)}$ indexed by the current source POI, $\operatorname{MLP}_{\mathrm{graph}}$ maps this propagated representation to the Transformer hidden dimension, and $\mathrm{LN}(\cdot)$ denotes layer normalization.

The motivation for multi-hop propagation is that an unseen direct transition may be supported by observed indirect paths.
We illustrate this with the two-hop case. For a current source POI $s$, the direct edge $(s,d)$ to a destination $d$ may be absent from the training graph, while the graph contains a path $s \rightarrow p' \rightarrow d$ through an intermediate POI $p'$.
In this case, two-hop propagation from $s$ exposes $d$ as a structurally plausible destination, providing soft transitive evidence for the missing transition $(s,d)$. More hops can cover more unseen-transition candidates, but they also introduce noisier and less source-specific evidence, making the graph token prone to over-smoothing~\cite{over_smoothing}.
We choose $N=2$ in our experiments, and Appendix~\ref{app:hops} analyzes this coverage-noise trade-off.

\subsection{User-History Revisit Scorer}
\label{sec:user_history_revisit_scorer}
Tail transitions can end at destinations that already appear in the user's observed history $\mathcal H_t$. What is new is the pairing of such a POI with the current source.
These are the \emph{user history revisits} introduced in Section~\ref{sec:introduction}.
RECAP treats each visited POI $d$ as the destination of a candidate transition $s_t\to d$ and scores this transition with a small network that reads statistics of the user's own history. It then uses these scores to adjust the core scores of the visited POIs.

\paragraph{Revisit Statistics.}
Let $\mathcal C_t=\{p_q\mid q\le t\}$ denote the set of POIs visited by the current user up to step $t$, including the current source $s_t$.
For each candidate $d\in\mathcal C_t$, we compute six statistics $\mathbf c_t(d)=[c_1,\dots,c_6]$ from the user's check-ins up to step $t$, listed in Table~\ref{tab:revisit-stats}.
Let $n_t(d)$ denote the number of the user's visits to $d$,  $\tau^{\mathrm{first}}_t(d)$ and $\tau^{\mathrm{last}}_t(d)$ denote the times of the first and the latest of these visits, and $m^u_t(a,d)$ denote the number of times the user moved from POI $a$ directly to $d$ within a trajectory.
The six statistics are
\begin{equation}
\label{eq:revisit-stats-def}
\begin{aligned}
c_1&=\frac{n_t(d)}{t}, &
c_2&=\tau^{\mathrm{last}}_t(d)-\tau^{\mathrm{first}}_t(d),\\
c_3&=\frac{m^u_t(s_t,d)}{\max\{1,\sum_{d'\in\mathcal C_t}m^u_t(s_t,d')\}}, &
c_4&=\lVert\operatorname{loc}(d)-\operatorname{loc}(s_t)\rVert_2,\\
c_5&=\textstyle\sum_{a:\,\operatorname{cat}(a)=\operatorname{cat}(s_t)}m^u_t(a,d), &
c_6&=\tau_t-\tau^{\mathrm{last}}_t(d),
\end{aligned}
\end{equation}
where $\tau_t$ is the time of the current check-in, and $c_2$ and $c_6$ are measured in days and hours. We omit $t$ and $d$ from $c_1,\dots,c_6$ for brevity.
The six statistics fall into two groups.
The first group ($c_1$ and $c_2$) describes the user's long-term relation with $d$.
The second group ($c_3$ to $c_6$) addresses Challenge~(2) by relating $d$ to the current context, namely where the user is now and when the current check-in happens, so the same POI can receive different scores at adjacent steps when the source or the time changes.

\begin{table}[t]
\caption{The six revisit statistics of a visited POI $d$ at step $t$.}
\label{tab:revisit-stats}
\small
\begin{tabular}{@{}p{1.3cm}cl@{}}
\toprule
Group & & Statistic \\
\midrule
\multirow{2}{=}{Long-term relation} & $c_1$ & Share of the user's check-ins at $d$ \\
 & $c_2$ & Days from the first to the latest visit to $d$ \\
\midrule
\multirow{4}{=}{Current context} & $c_3$ & Share of the user's moves from $s_t$ that go to $d$ \\
 & $c_4$ & Distance from $s_t$ to $d$ \\
 & $c_5$ & Moves into $d$ from POIs of the category of $s_t$ \\
 & $c_6$ & Hours since the latest visit to $d$ \\
\bottomrule
\end{tabular}
\end{table}

\paragraph{Revisit Scorer.}
A scorer $f$ maps the statistics of each visited POI to a scalar score. A softmax over the visited POIs turns the scores into a revisit distribution, the probability that the user goes to $d$ next if the next POI is one of the visited POIs:
\begin{equation}
\label{eq:revisit-prob}
\pi^{\mathrm{rev}}_t(d)=\frac{\exp f(\mathbf c_t(d))}{\sum_{d'\in\mathcal C_t}\exp f(\mathbf c_t(d'))},\qquad d\in\mathcal C_t.
\end{equation}
Tail transitions have few or no training cases. Parameters tied to a specific POI or transition receive little training signal for them.
We therefore let $f$ take only the six statistics as input and share it across all users and POIs, so that what $f$ learns from frequent revisits also applies to rare ones.
We implement $f$ as a three-layer MLP with GELU activations, to learn how the six statistics jointly determine whether the user returns to $d$.

\paragraph{Combining with the Backbone.}
RECAP adjusts the core score of each visited POI and leaves the other POIs unchanged:
\begin{equation}
\label{eq:final-logit}
\ell_t(d)=
\begin{cases}
\ell^{\mathrm{core}}_t(d)+\lambda\log\big(1+K_t\,\pi^{\mathrm{rev}}_t(d)\big), & d\in\mathcal C_t,\\
\ell^{\mathrm{core}}_t(d), & d\in\mathcal P\setminus\mathcal C_t.
\end{cases}
\end{equation}
Here, $K_t=|\mathcal C_t|$ is the number of POIs that the user has visited, which differs across users.
Since $\pi^{\mathrm{rev}}_t$ is spread over these $K_t$ POIs, its values shrink as $K_t$ grows.
Multiplying by $K_t$ rescales them to the same scale for all users: $K_t\,\pi^{\mathrm{rev}}_t(d)$ equals one for a POI at the average share $1/K_t$, exceeds one for a POI that the scorer prefers, and is close to zero for a POI that the scorer considers unlikely.
Adding one inside the logarithm keeps the adjustment non-negative, so a visited POI that the scorer considers unlikely keeps about its core score.
The hyperparameter $\lambda$ sets the weight of the revisit scorer against the backbone.

\subsection{Two-Stage Training}
\label{sec:two_stage_training}
Tail transitions have few training cases, so the revisit scorer has to learn mainly from frequent transitions whose target is a POI that the user has visited.
Let $\mathcal D_{\mathrm{train}}$ denote the set of prediction cases in the training data.
Suppose the backbone and the revisit scorer were trained jointly with the cross-entropy of the final score $\ell_t$ in Equation~\eqref{eq:final-logit}.
For a visited POI $d$, the gradient of this loss with respect to the adjustment of $d$ is $\pi_t(d)-\mathbf 1[d=y_t]$, where $\pi_t(d)$ is the softmax probability of $d$ under $\ell_t$, and $\mathbf 1[d=y_t]$ equals one if $d$ is the target, and zero otherwise.
This gradient becomes smaller as the backbone ranks the target higher, even when the scorer ranks the target low.
The frequent transitions that the backbone already ranks well thus give the scorer little training signal.
RECAP therefore trains the scorer after the backbone, with the backbone frozen and a separate loss on its own revisit distribution $\pi^{\mathrm{rev}}_t$.

In the first stage, we train the backbone, including the graph token, with the cross-entropy over all POIs on all training cases:
\begin{equation}
\label{eq:core-loss}
\mathcal L_{\mathrm{core}}=-\frac{1}{|\mathcal D_{\mathrm{train}}|}\sum_{t\in\mathcal D_{\mathrm{train}}}\log\frac{\exp(\ell^{\mathrm{core}}_t(y_t))}{\sum_{d\in\mathcal P}\exp(\ell^{\mathrm{core}}_t(d))}.
\end{equation}
In the second stage, we freeze all backbone weights and train only the scorer $f$ on the subset $\mathcal D^{\mathrm{rev}}_{\mathrm{train}}=\{t\in\mathcal D_{\mathrm{train}}\mid y_t\in\mathcal C_t\}$ of training cases whose target is in the user's history:
\begin{equation}
\label{eq:rev-loss}
\mathcal L_{\mathrm{rev}}=-\frac{1}{|\mathcal D^{\mathrm{rev}}_{\mathrm{train}}|}\sum_{t\in\mathcal D^{\mathrm{rev}}_{\mathrm{train}}}\log \pi^{\mathrm{rev}}_t(y_t).
\end{equation}
The gradient of $\mathcal L_{\mathrm{rev}}$ with respect to the score of the target is $\pi^{\mathrm{rev}}_t(y_t)-1$.
It stays large whenever the scorer ranks the target low among the visited POIs, however well the backbone fits the same case.
Frequent transitions whose target is a visited POI therefore also train the scorer in this stage.
Since $\mathcal L_{\mathrm{rev}}$ does not depend on the backbone, the two stages need not run in sequence and can also be trained in parallel. What matters is that the scorer is not trained through the final score.
At inference, RECAP computes the core scores and the revisit distribution in a single forward pass and ranks all POIs by the final score $\ell_t$ in Equation~\eqref{eq:final-logit}.

\section{Experiments}
\label{sec:experiments}

\paragraph{Datasets and Baselines}
We evaluate our model on three public real-world datasets: NYC~\cite{nyctky}, TKY~\cite{nyctky}, and CA~\cite{ca}.
Appendix~\ref{app:datasets} reports the statistics of the three datasets, and Appendix~\ref{app:tail-definition-sensitivity} reports their transition-frequency distributions.
We compare RECAP with representative baselines, including (1)~\textbf{sequential models} DeepMove~\cite{deepmove}, Flashback~\cite{flashback}, ROTAN~\cite{rotan}, REPLAY~\cite{replay}, DPRL~\cite{dprl}, and K1-POI~\cite{k1poi}, (2)~\textbf{graph and hypergraph models} GETNext~\cite{getnext}, STHGCN~\cite{sthgcn}, DCHL~\cite{dchl}, and MSAHG~\cite{msahg}, (3)~\textbf{LLM-based methods} LLM4POI~\cite{llm4poi}, AgentMove~\cite{agentmove}, and GNPR-SID~\cite{gnpr_sid}, and (4)~\textbf{long-tail targeted methods} MELT~\cite{melt}, LoTNext~\cite{lotnext}, DePOI~\cite{depoi}, and ALOHA~\cite{aloha}~(see Appendix~\ref{app:compared_methods} for details). We re-implement or adapt all baselines and evaluate them under the same split, preprocessing, and metrics.
All the implementation details are described in Appendix~\ref{app:implementation-details}.

\begin{table*}[!tbp]
\caption{Performance comparison in HR and NDCG on three datasets. For compactness, table headers abbreviate HR@K as H@K and NDCG@K as N@K. The final row reports RECAP's relative gain over the best baseline in each column.}
\label{tab:main_results}
\small
\begin{tabular}{lc@{\;}c@{\;}c@{\;}c@{\;}cc@{\;}c@{\;}c@{\;}c@{\;}cc@{\;}c@{\;}c@{\;}c@{\;}c}
\toprule
\multirow{2}{*}[-1ex]{Method} & \multicolumn{5}{c}{NYC} & \multicolumn{5}{c}{TKY} & \multicolumn{5}{c}{CA} \\
\cmidrule(lr){2-6}
\cmidrule(lr){7-11}
\cmidrule(lr){12-16}
& H@1 & H@10 & H@20 & N@10 & N@20 & H@1 & H@10 & H@20 & N@10 & N@20 & H@1 & H@10 & H@20 & N@10 & N@20 \\
\midrule
DeepMove$\mathrm{_{\tiny(WWW'18)}}$ & 0.1526 & 0.4312 & 0.4988 & 0.2830 & 0.3002 & 0.1809 & 0.4716 & 0.5469 & 0.3165 & 0.3356 & 0.1026 & 0.2901 & 0.3543 & 0.1885 & 0.2047 \\
Flashback$\mathrm{_{\tiny(IJCAI'20)}}$ & 0.1789 & 0.5191 & 0.5924 & 0.3377 & 0.3564 & 0.1682 & 0.4558 & 0.5273 & 0.3031 & 0.3212 & 0.1233 & 0.3275 & 0.3889 & 0.2187 & 0.2342 \\
ROTAN$\mathrm{_{\tiny(KDD'24)}}$ & 0.2185 & 0.6172 & 0.6924 & 0.4064 & 0.4256 & 0.2467 & 0.5965 & 0.6718 & 0.4113 & 0.4304 & 0.1482 & 0.4028 & 0.4780 & 0.2664 & 0.2854 \\
REPLAY$\mathrm{_{\tiny(TMC'25)}}$ & 0.1776 & 0.5123 & 0.5821 & 0.3347 & 0.3525 & 0.1786 & 0.4715 & 0.5411 & 0.3164 & 0.3341 & 0.1188 & 0.3212 & 0.3821 & 0.2133 & 0.2287 \\
DPRL$\mathrm{_{\tiny(IJCAI'25)}}$ & 0.1794 & 0.5159 & 0.5892 & 0.3378 & 0.3565 & 0.2087 & 0.5417 & 0.6157 & 0.3652 & 0.3841 & 0.1336 & 0.3789 & 0.4547 & 0.2463 & 0.2655 \\
K1-POI$\mathrm{_{\tiny(KDD'26)}}$ & 0.2289 & \underline{0.6364} & 0.7122 & \underline{0.4213} & \underline{0.4406} & 0.2463 & \underline{0.6101} & \underline{0.6878} & \underline{0.4174} & \underline{0.4372} & \underline{0.1596} & \underline{0.4382} & \underline{0.5242} & \underline{0.2883} & \underline{0.3101} \\
\midrule
GETNext$\mathrm{_{\tiny(SIGIR'22)}}$ & 0.1766 & 0.4690 & 0.5413 & 0.3157 & 0.3341 & 0.1503 & 0.3350 & 0.3832 & 0.2423 & 0.2546 & 0.1192 & 0.2849 & 0.3134 & 0.2003 & 0.2075 \\
STHGCN$\mathrm{_{\tiny(SIGIR'23)}}$ & 0.2290 & 0.6255 & 0.6968 & 0.4169 & 0.4351 & \underline{0.2485} & 0.5915 & 0.6660 & 0.4106 & 0.4296 & 0.1461 & 0.4064 & 0.4864 & 0.2660 & 0.2862 \\
DCHL$\mathrm{_{\tiny(SIGIR'24)}}$ & 0.1841 & 0.6199 & \underline{0.7216} & 0.3859 & 0.4119 & 0.1301 & 0.4970 & 0.5919 & 0.2954 & 0.3195 & 0.1025 & 0.3745 & 0.4616 & 0.2272 & 0.2492 \\
MSAHG$\mathrm{_{\tiny(AAAI'26)}}$ & 0.1851 & 0.6036 & 0.7071 & 0.3751 & 0.4015 & 0.1400 & 0.4964 & 0.5963 & 0.3009 & 0.3263 & 0.1118 & 0.3542 & 0.4228 & 0.2244 & 0.2418 \\
\midrule
LLM4POI$\mathrm{_{\tiny(SIGIR'24)}}$ & 0.2077 & 0.5995 & 0.6461 & 0.3942 & 0.4062 & 0.1954 & 0.5255 & 0.5757 & 0.3518 & 0.3647 & 0.1253 & 0.3415 & 0.3722 & 0.2301 & 0.2380 \\
AgentMove$\mathrm{_{\tiny(NAACL'25)}}$ & 0.1626 & 0.5499 & 0.5828 & 0.3463 & 0.3548 & 0.1266 & 0.4486 & 0.4895 & 0.2769 & 0.2875 & 0.0907 & 0.2865 & 0.3077 & 0.1824 & 0.1879 \\
GNPR-SID$\mathrm{_{\tiny(KDD'25)}}$ & 0.2230 & 0.6134 & 0.6558 & 0.4115 & 0.4224 & 0.2108 & 0.5240 & 0.5897 & 0.3607 & 0.3774 & 0.1548 & 0.3725 & 0.4200 & 0.2575 & 0.2696 \\
\midrule
MELT$\mathrm{_{\tiny(SIGIR'23)}}$ & 0.2121 & 0.6308 & 0.7148 & 0.4068 & 0.4283 & 0.2253 & 0.5979 & 0.6771 & 0.4002 & 0.4203 & 0.1280 & 0.3818 & 0.4616 & 0.2457 & 0.2660 \\
LoTNext$\mathrm{_{\tiny(NeurIPS'24)}}$ & 0.1956 & 0.5403 & 0.6144 & 0.3591 & 0.3780 & 0.2141 & 0.5541 & 0.6252 & 0.3763 & 0.3943 & 0.1380 & 0.3662 & 0.4368 & 0.2442 & 0.2621 \\
DePOI$\mathrm{_{\tiny(SIGIR'25)}}$ & 0.1783 & 0.5950 & 0.6716 & 0.3723 & 0.3918 & 0.1518 & 0.5112 & 0.5983 & 0.3192 & 0.3413 & 0.0828 & 0.3511 & 0.4469 & 0.2029 & 0.2272 \\
ALOHA$\mathrm{_{\tiny(WWW'26)}}$ & \underline{0.2298} & 0.6241 & 0.7003 & 0.4161 & 0.4356 & 0.2258 & 0.5654 & 0.6480 & 0.3844 & 0.4054 & 0.1423 & 0.4067 & 0.4878 & 0.2639 & 0.2845 \\
\midrule
RECAP\rlap{$^{*}$}$\mathrm{_{\tiny(Ours)}}$  & \textbf{0.2837} & \textbf{0.7402} & \textbf{0.8025} & \textbf{0.5030} & \textbf{0.5190} & \textbf{0.2751} & \textbf{0.6727} & \textbf{0.7485} & \textbf{0.4643} & \textbf{0.4836} & \textbf{0.1963} & \textbf{0.4829} & \textbf{0.5615} & \textbf{0.3313} & \textbf{0.3512} \\
\emph{Rel. Gain} & $+23.5\%$ & $+16.3\%$ & $+11.2\%$ & $+19.4\%$ & $+17.8\%$ & $+10.7\%$ & $+10.3\%$ & $+8.8\%$ & $+11.2\%$ & $+10.6\%$ & $+23.0\%$ & $+10.2\%$ & $+7.1\%$ & $+14.9\%$ & $+13.3\%$ \\
\bottomrule
\multicolumn{16}{@{}l}{\footnotesize $^{*}$ The gain of RECAP is significant in every column under Welch's $t$-test over three seeds ($p<0.01$).} \\
\end{tabular}
\end{table*}

\paragraph{Evaluation Metrics}
We use Hit Ratio (HR@$K$) and Normalized Discounted Cumulative Gain (NDCG@$K$), which measure top-$K$ ranking quality. We report HR@$K$ at $K=1$, $10$, and $20$ and NDCG@$K$ at $K=10$ and $20$ to capture both fine-grained accuracy and broader recall.
In training and testing, every check-in of a trajectory except the first one is a prediction case, and all methods are evaluated on the same test cases. Following~\cite{flashback,k1poi}, we evaluate every such check-in rather than only the last check-in of each trajectory or user, so that the evaluation also covers check-ins with short histories.
We report the mean value over three different seeds.

\subsection{Experimental Results}

\begin{table}[!t]
\centering
\caption{Ablation results.}
\label{tab:ablation}
\small
\resizebox{\columnwidth}{!}{%
\begin{tabular}{lc@{\;}cc@{\;}cc@{\;}c}
\toprule
\multirow{2}{*}[-1ex]{Variant} & \multicolumn{2}{c}{NYC} & \multicolumn{2}{c}{TKY} & \multicolumn{2}{c}{CA} \\
\cmidrule(lr){2-3}
\cmidrule(lr){4-5}
\cmidrule(lr){6-7}
& H@1 & H@20 & H@1 & H@20 & H@1 & H@20 \\
\midrule
Backbone & 0.2086 & 0.7170 & 0.2255 & 0.6733 & 0.1429 & 0.4793 \\
\;+ Graph token & 0.2206 & 0.7175 & 0.2292 & 0.6852 & 0.1505 & 0.5115 \\
\;+ Revisit scorer (\textbf{RECAP}) & \textbf{0.2837} & \textbf{0.8025} & \textbf{0.2751} & \textbf{0.7485} & \textbf{0.1963} & \textbf{0.5615} \\
\midrule
Joint training & 0.2620 & 0.7852 & 0.2707 & 0.7326 & 0.1939 & 0.5355 \\
\midrule
w/o long-term relation & 0.2700 & 0.7703 & 0.2678 & 0.7375 & 0.1814 & 0.5563 \\
w/o current context & 0.2262 & 0.7251 & 0.2351 & 0.6978 & 0.1592 & 0.5308 \\
Linear scorer & 0.2588 & 0.7893 & 0.2461 & 0.7157 & 0.1668 & 0.5330 \\
\bottomrule
\end{tabular}%
}
\end{table}

\begin{table*}[!t]
\caption{Results on tail and unseen transitions. Tail transitions appear at most once in the training data, and unseen transitions never appear in it. The final row reports RECAP's relative gain over the best baseline in each column.}
\label{tab:full_head_tail_results}
\small
\setlength{\tabcolsep}{3.4pt}
\begin{tabular}{lc@{\,}c@{\,}c@{\,}c@{\,}c@{\,}cc@{\,}c@{\,}c@{\,}c@{\,}c@{\,}cc@{\,}c@{\,}c@{\,}c@{\,}c@{\,}c}
\toprule
\multirow{3}{*}[-1.4ex]{Method} & \multicolumn{6}{c}{NYC} & \multicolumn{6}{c}{TKY} & \multicolumn{6}{c}{CA} \\
\cmidrule(lr){2-7}
\cmidrule(lr){8-13}
\cmidrule(lr){14-19}
& \multicolumn{3}{c}{Tail} & \multicolumn{3}{c}{Unseen} & \multicolumn{3}{c}{Tail} & \multicolumn{3}{c}{Unseen} & \multicolumn{3}{c}{Tail} & \multicolumn{3}{c}{Unseen} \\
\cmidrule(lr){2-4}\cmidrule(lr){5-7}\cmidrule(lr){8-10}\cmidrule(lr){11-13}\cmidrule(lr){14-16}\cmidrule(lr){17-19}
& H@1 & H@20 & N@20 & H@1 & H@20 & N@20 & H@1 & H@20 & N@20 & H@1 & H@20 & N@20 & H@1 & H@20 & N@20 & H@1 & H@20 & N@20 \\
\midrule
DeepMove & 0.0512 & 0.3058 & 0.1490 & 0.0393 & 0.2400 & 0.1151 & 0.0393 & 0.2601 & 0.1215 & 0.0281 & 0.2055 & 0.0933 & 0.0310 & 0.1982 & 0.0940 & 0.0226 & 0.1502 & 0.0695 \\
Flashback & 0.0625 & 0.4143 & 0.1982 & 0.0434 & 0.3268 & 0.1482 & 0.0417 & 0.2792 & 0.1312 & 0.0318 & 0.2306 & 0.1046 & 0.0460 & 0.2626 & 0.1301 & 0.0306 & 0.2112 & 0.0979 \\
ROTAN & 0.0995 & 0.5449 & 0.2760 & 0.0771 & 0.4579 & 0.2230 & 0.0683 & 0.4406 & 0.2110 & 0.0516 & 0.3745 & 0.1719 & 0.0595 & 0.3450 & 0.1705 & 0.0417 & 0.2772 & 0.1301 \\
REPLAY & 0.0654 & 0.4017 & 0.1976 & 0.0476 & 0.3196 & 0.1511 & 0.0487 & 0.2935 & 0.1425 & 0.0374 & 0.2448 & 0.1154 & 0.0484 & 0.2590 & 0.1303 & 0.0369 & 0.2110 & 0.1029 \\
DPRL & 0.0728 & 0.4069 & 0.2038 & 0.0536 & 0.3212 & 0.1555 & 0.0572 & 0.3703 & 0.1772 & 0.0429 & 0.3096 & 0.1431 & 0.0471 & 0.3107 & 0.1482 & 0.0340 & 0.2415 & 0.1108 \\
K1-POI & 0.1001 & 0.5735 & 0.2896 & 0.0785 & 0.4929 & 0.2396 & 0.0710 & 0.4596 & 0.2208 & 0.0545 & 0.3918 & 0.1817 & 0.0641 & \underline{0.3926} & \underline{0.1913} & 0.0438 & \underline{0.3219} & 0.1479 \\
\midrule
GETNext & 0.0310 & 0.3312 & 0.1457 & 0.0098 & 0.2323 & 0.0872 & 0.0066 & 0.1101 & 0.0449 & 0.0002 & 0.0677 & 0.0216 & 0.0118 & 0.1384 & 0.0623 & 0.0006 & 0.0422 & 0.0153 \\
STHGCN & 0.1105 & 0.5525 & 0.2892 & 0.0945 & 0.4772 & 0.2463 & 0.0710 & 0.4324 & 0.2111 & 0.0548 & 0.3670 & 0.1725 & 0.0687 & 0.3570 & 0.1798 & 0.0573 & 0.3081 & 0.1523 \\
DCHL & 0.1159 & \underline{0.6025} & 0.3127 & 0.0970 & \underline{0.5331} & 0.2719 & 0.0641 & 0.3955 & 0.1923 & 0.0563 & 0.3520 & 0.1702 & 0.0632 & 0.3600 & 0.1779 & 0.0547 & 0.3160 & 0.1541 \\
MSAHG & 0.1145 & 0.5901 & 0.3030 & 0.1011 & 0.5304 & 0.2704 & 0.0789 & 0.4222 & 0.2113 & 0.0712 & 0.3851 & 0.1918 & 0.0710 & 0.3450 & 0.1792 & 0.0618 & 0.3160 & 0.1607 \\
\midrule
LLM4POI & \underline{0.1528} & 0.5511 & \underline{0.3303} & \textbf{0.1348} & 0.5054 & \underline{0.2984} & \underline{0.0959} & 0.4036 & 0.2255 & \textbf{0.0874} & 0.3738 & \underline{0.2069} & \underline{0.0898} & 0.3043 & 0.1842 & \underline{0.0840} & 0.2822 & \underline{0.1699} \\
AgentMove & 0.1240 & 0.4932 & 0.2920 & 0.1111 & 0.4514 & 0.2641 & 0.0725 & 0.3395 & 0.1853 & 0.0663 & 0.3141 & 0.1703 & 0.0616 & 0.2480 & 0.1419 & 0.0550 & 0.2298 & 0.1292 \\
GNPR-SID & 0.1489 & 0.5524 & 0.3281 & 0.1329 & 0.5032 & 0.2953 & 0.0803 & 0.3674 & 0.1961 & 0.0686 & 0.3257 & 0.1709 & 0.0883 & 0.3122 & 0.1819 & 0.0750 & 0.2793 & 0.1594 \\
\midrule
MELT & 0.1131 & 0.5828 & 0.2990 & 0.0985 & 0.5115 & 0.2607 & 0.0800 & \underline{0.4681} & \underline{0.2305} & 0.0667 & \underline{0.4121} & 0.1984 & 0.0685 & 0.3606 & 0.1825 & 0.0589 & 0.3202 & 0.1589 \\
LoTNext & 0.0721 & 0.4353 & 0.2141 & 0.0491 & 0.3427 & 0.1596 & 0.0720 & 0.3831 & 0.1931 & 0.0553 & 0.3215 & 0.1566 & 0.0498 & 0.2975 & 0.1454 & 0.0303 & 0.2320 & 0.1049 \\
DePOI & 0.0867 & 0.5138 & 0.2567 & 0.0688 & 0.4283 & 0.2078 & 0.0470 & 0.3333 & 0.1534 & 0.0358 & 0.2763 & 0.1242 & 0.0423 & 0.3157 & 0.1418 & 0.0319 & 0.2556 & 0.1116 \\
ALOHA & 0.1147 & 0.5577 & 0.2930 & 0.0990 & 0.4824 & 0.2511 & 0.0732 & 0.4276 & 0.2112 & 0.0585 & 0.3661 & 0.1752 & 0.0693 & 0.3593 & 0.1811 & 0.0576 & 0.3108 & 0.1536 \\
\midrule
RECAP\rlap{$^{*}$}$\mathrm{_{\tiny(Ours)}}$ & \textbf{0.1609} & \textbf{0.7058} & \textbf{0.3913} & \underline{0.1338} & \textbf{0.6455} & \textbf{0.3454} & \textbf{0.1026} & \textbf{0.5771} & \textbf{0.2912} & \underline{0.0810} & \textbf{0.5265} & \textbf{0.2529} & \textbf{0.1130} & \textbf{0.4742} & \textbf{0.2615} & \textbf{0.0948} & \textbf{0.4360} & \textbf{0.2320} \\
\emph{Rel. Gain} & $+5.3\%$ & $+17.1\%$ & $+18.5\%$ & $-0.7\%$ & $+21.1\%$ & $+15.8\%$ & $+7.0\%$ & $+23.3\%$ & $+26.3\%$ & $-7.3\%$ & $+27.8\%$ & $+22.2\%$ & $+25.8\%$ & $+20.8\%$ & $+36.7\%$ & $+12.9\%$ & $+35.4\%$ & $+36.6\%$ \\
\bottomrule
\multicolumn{19}{@{}l}{\footnotesize $^{*}$ In every column where RECAP is the best, its gain over the best baseline is significant under Welch's $t$-test over three seeds ($p<0.01$).} \\
\end{tabular}
\end{table*}

\paragraph{Overall Performance.}
Table~\ref{tab:main_results} reports the overall comparison results. RECAP achieves the best result (in boldface) on all three datasets and all evaluation metrics. Compared with the best baseline (underlined) in each column, RECAP improves by 7.1\% to 23.5\%. The gain is largest on HR@1, with 23.5\%, 10.7\%, and 23.0\% on NYC, TKY, and CA, respectively, and becomes smaller as $K$ grows, showing that RECAP improves most on the top-ranked prediction. Across the three seeds, the standard deviation of RECAP is at most 0.0024, and its smallest gain over the best baseline, 0.0266 in HR@1 on TKY, is more than ten times this value.

Among the baselines, K1-POI is the strongest competitor and is the best baseline in 12 of the 15 columns. K1-POI predicts the next POI from only the current POI, so its training amounts to memorizing source--destination transitions, which benefits frequent transitions more than unseen ones (see Table~\ref{tab:full_head_tail_results}). Graph-based methods such as MSAHG and DCHL are weaker on HR@1 but stronger on HR@20, and DCHL has the best baseline HR@20 on NYC, suggesting that graph structures help retrieve a broader candidate set. LLM-based methods use the world knowledge of the language model to exploit the revisit signals in the input history, and GNPR-SID, the strongest of them, is close to the best baseline on HR@1 on NYC and CA. However, instead of scoring every POI in the dataset, they generate candidates token by token, and these candidates concentrate on POIs that appear in the input check-ins, so they recall fewer plausible POIs and fall further behind on HR@20. Moreover, compared with POI-level long-tail methods, RECAP obtains consistently better results across all datasets, showing that modeling the transition level long-tail problem is more effective.

\paragraph{Ablation Study.}
\label{sec:ablation}

Table~\ref{tab:ablation} adds the components of RECAP to the backbone one at a time. The graph token improves the backbone on all three datasets, especially in HR@20 on CA, and the revisit scorer brings the largest improvement.

``Joint training'' trains the backbone and the scorer jointly instead of in two stages, which lowers the performance on all three datasets. This is consistent with Section~\ref{sec:two_stage_training}: the training cases that the backbone already ranks well give the scorer little training signal.

We further examine the design of the revisit scorer. Removing either group of statistics in Table~\ref{tab:revisit-stats} (``w/o long-term relation'' and ``w/o current context'') lowers the performance on all three datasets. Removing the current context causes the larger drop and leaves the scorer little gain over the backbone, so scoring the same visited POI differently as the source and the time change is the main source of the gain. Replacing the scorer with a linear combination of the six statistics (``Linear scorer'') also lowers the performance, showing that the statistics need to be combined nonlinearly.

\begin{figure*}[!t]
\centering
\includegraphics[width=0.96\textwidth,trim={0 0 0 3.5bp},clip]{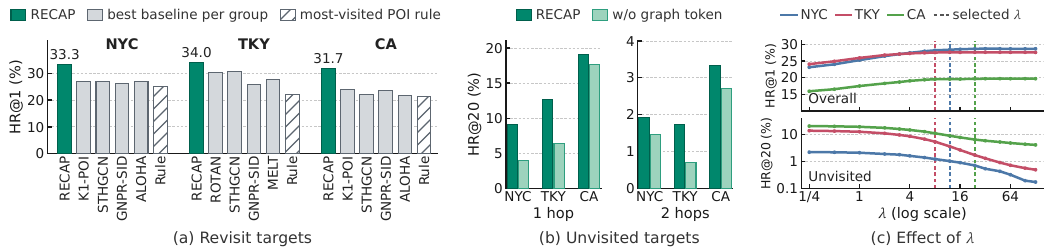}
\caption{(a) HR@1 on revisit targets. (b) HR@20 on unvisited targets that are reachable from the current POI within one or two hops of the transition graph. (c) Overall HR@1 and HR@20 on unvisited targets under different $\lambda$.}
\label{fig:evidence}
\end{figure*}

\paragraph{Performance on Long-Tail Transitions.}
Table~\ref{tab:full_head_tail_results} reports the results on tail transitions, which appear at most once in training, and unseen transitions, which never appear.
RECAP achieves the best tail HR@1 and HR@20 on all three datasets, and improves tail HR@20 over the best baseline by 17.1\%, 23.3\%, and 20.8\%. On unseen transitions, the HR@20 gains increase to 21.1\%, 27.8\%, and 35.4\%. On CA, the sparsest of the three datasets, RECAP achieves the best results on both tail and unseen transitions in all metrics.

LLM-based methods are relatively strong on HR@1 for these transitions. Most tail transitions end at POIs the user has visited before, which LLM-based methods can find in the input history, whereas sequential and graph models have few or no training samples for these transitions. However, their candidates, generated token by token, concentrate on POIs in the input check-ins, which limits their recall of other plausible destinations, so RECAP outperforms them on HR@20 by a large margin, e.g., 0.5265 vs. 0.3738 for LLM4POI on unseen transitions in TKY.

\begin{figure*}[!t]
\centering
\includegraphics[width=\textwidth,trim={0 1.5bp 0 1.5bp},clip]{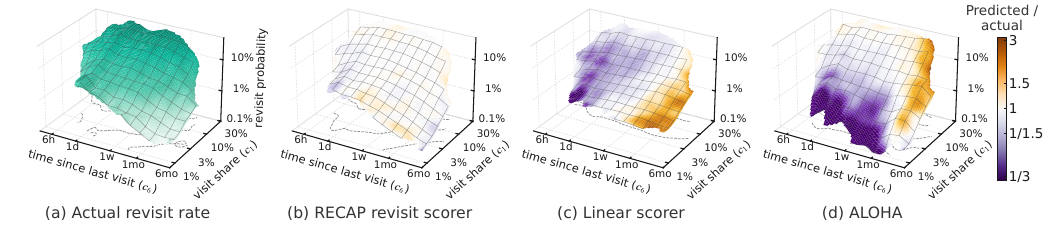}
\caption{Revisit probability of visited POIs on NYC against the user's check-in share at the POI ($c_1$) and the hours since the latest visit ($c_6$). (a) Actual revisit rate. (b) Mean $\pi^{\mathrm{rev}}$ of the revisit scorer. (c) Mean $\pi^{\mathrm{rev}}$ of a linear scorer. (d) Mean probability of ALOHA, renormalized over the visited POIs. Colors in (b)--(d) show the ratio of the predicted to the actual rate.}
\label{fig:revisit-surface}
\end{figure*}

\subsection{Analysis of Revisit and Transitive Evidence}
\label{sec:evidence-analysis}

We split the test samples by whether the target is a POI that the user has visited before.  The targets visited before (``revisit targets'' hereafter) account for most test samples and most tail transitions. Figure~\ref{fig:evidence}(a) shows that RECAP achieves the best HR@1 on revisit targets on all three datasets. Meanwhile, simply predicting the user's most visited POI (``Rule'' in the figure) stays well below RECAP. This shows that knowing which POIs the user has visited is not enough. STHGCN and ALOHA also link each prediction to the user's earlier trajectories in the training data, yet stay below RECAP (Table~\ref{tab:main_results}). The visited POIs need to be scored under the current context (e.g., the current POI and the time since the last visit) as we did.

Figure~\ref{fig:revisit-surface} shows what the revisit scorer learns on NYC.
For the test samples whose target the user has visited, we group the visited POIs by the user's check-in share at the POI ($c_1$) and the hours since the latest visit ($c_6$), and compare the actual revisit rate with the mean revisit probability $\pi^{\mathrm{rev}}$. Figure~\ref{fig:revisit-surface}(a) shows that the actual rate grows with $c_1$, peaks about one day after the latest visit, and then falls, so the surface is curved. The revisit scorer of RECAP follows this surface, and its probability stays close to the actual rate (Figure~\ref{fig:revisit-surface}(b)). A linear combination of the same statistics cannot bend the surface (Figure~\ref{fig:revisit-surface}(c)): it underestimates POIs visited within the last day and overestimates POIs not visited for weeks, which is consistent with its drop in Table~\ref{tab:ablation}. ALOHA, the strongest baseline on revisit targets on NYC, deviates by a similar factor (Figure~\ref{fig:revisit-surface}(d)): it underestimates POIs with a small check-in share that the user visited within the last month and overestimates POIs with a larger share that the user has not visited for over a month. Appendix~\ref{app:error-analysis} further analyzes the errors of RECAP on visited targets.

For targets that the user has not visited, the graph token supplies transitive evidence. Figure~\ref{fig:evidence}(b) shows the samples whose target is reachable from the current POI within one or two hops of the transition graph. The graph token improves HR@20 on these targets on all three datasets. Reaching farther targets would require more hops, but three hops enlarge the candidate set several times and bring in many unrelated POIs that dilute the transition signal. Appendix~\ref{app:hops} analyzes this trade-off.

In Equation~\eqref{eq:final-logit}, $\lambda$ scales the bonus that the revisit scorer adds to the core score of each POI the user has visited. A larger $\lambda$ ranks visited POIs further ahead of unvisited ones. Figure~\ref{fig:evidence}(c) shows that as $\lambda$ grows, the overall HR@1 rises and then levels off, while HR@20 on targets that the user has not visited keeps falling, because visited POIs take more positions in the top 20. We select $\lambda$ on the validation set. The choice of $\lambda$ thus depends on the needs of the service. A service that values accuracy can use the $\lambda$ selected on validation, while a service that values recommending new places can use a smaller $\lambda$, which also lets the graph token recall more of these places, at some cost in overall accuracy. For example, on CA, $\lambda=1$ raises HR@20 on unvisited targets from 0.066 to 0.196 and lowers the overall HR@1 from 0.196 to 0.175, and the gain from the graph token on this HR@20 grows from 0.021 to 0.068.

\begin{figure}[!t]
\centering
\includegraphics[width=0.9\columnwidth,trim={0 5.5bp 0 1.5bp},clip]{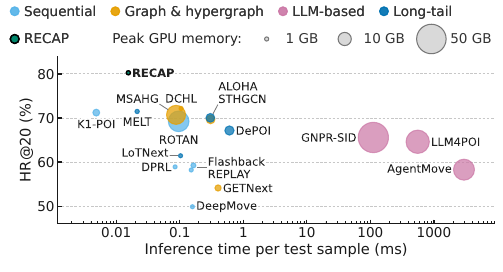}
\caption{Accuracy and efficiency comparison on NYC. Inference time is averaged over test samples and shown on a log scale, and marker area indicates peak GPU memory.}
\label{fig:efficiency}
\end{figure}

\subsection{Efficiency Analysis}
\label{sec:efficiency}

Figure~\ref{fig:efficiency} compares the performance and efficiency of RECAP and the baselines on NYC. RECAP lies at the upper left on all three datasets, achieving the best balance between performance and efficiency. The LLM-based baselines take about 0.1 to 3.4 seconds per test sample and 25 to 53 GB of GPU memory, which makes them difficult to deploy in online Web recommendation services. Appendix~\ref{app:efficiency-analysis} reports the detailed comparison on all three datasets, including the training time, test time, and GPU memory of all methods.

\begin{figure}[!t]
    \centering
    \includegraphics[width=0.9\columnwidth]{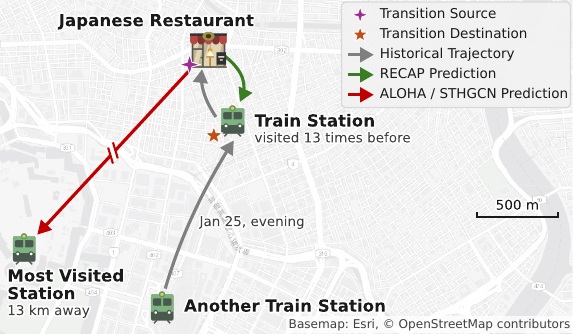}
    \caption{A prediction example for an unseen transition.}
    \label{fig:case_map}
\end{figure}

\subsection{Case Study}
Figure~\ref{fig:case_map} shows a prediction on an unseen transition on TKY. The user first arrives at a \texttt{Train Station} and then checks in at a \texttt{Japanese Restaurant} nearby. The ground-truth next POI is the same station. The station is a popular POI with 706 training check-ins, yet the transition \texttt{Japanese Restaurant} $\rightarrow$ \texttt{Train Station} never appears in the training data. LoTNext predicts that the user stays at the restaurant, and ALOHA and STHGCN tend to predict another station 13 km away, which the user visits most often. RECAP's revisit scorer ranks the station first among the 15 POIs this user has visited, since the user has visited it 13 times and passed through it a few hours earlier, and RECAP places it at rank 1.

\section{Conclusion}
\label{sec:conclusion}
As mobility prediction moves toward increasingly heavier architectures, from hypergraphs to LLM-based generators, this paper points to a simpler fault line: \emph{transitions}. Many errors arise from long-tailed source--destination compositions, even when the destination is frequently seen in the training data. RECAP shows that a lightweight Transformer given transition evidence can outperform more complex models. Future work can consider more challenging yet realistic settings, such as cold-start prediction for new POIs, new users, or even new cities that never appear during training. These settings make the problem of long-tailed transitions more severe and require stronger out-of-distribution generalization abilities.

\bibliographystyle{ACM-Reference-Format}
\bibliography{references}

\appendix

\section{Experimental Setup}
\label{app:experimental-setup}

\subsection{Datasets}
\label{app:datasets}

Following \citet{getnext}, we sort the check-ins of the NYC, TKY~\cite{nyctky}, and CA~\cite{ca} datasets in chronological order and split each into training, validation, and test sets with a ratio of 8:1:1, and Table~\ref{tab:dataset_statistics} reports their statistics.

\begin{table}[!h]
\centering
\caption{Dataset statistics. Tail and head percentages are computed over unique test source--destination pairs.}
\label{tab:dataset_statistics}
\small
\begin{tabular}{lrrr}
\toprule
 & NYC & TKY & CA \\
\midrule
\#Users & 1,074 & 2,281 & 3,652 \\
\#POIs & 5,069 & 7,836 & 9,756 \\
\#Categories & 321 & 291 & 298 \\
\#Check-ins & 134,481 & 424,585 & 221,359 \\
\#Test transitions & 10,309 & 35,329 & 10,434 \\
Tail & 77.9\% & 65.4\% & 82.1\% \\
Head & 22.1\% & 34.6\% & 17.9\% \\
\bottomrule
\end{tabular}
\end{table}

\subsection{Details of Compared Methods}
\label{app:compared_methods}

We compare RECAP with the four groups of methods in Table~\ref{tab:main_results}, and Section~\ref{sec:related} describes most of them. Among the others, K1-POI~\cite{k1poi} recalls candidates from the user and the last check-in with a self-attention model and reranks them with temporal and spatial priors, AgentMove~\cite{agentmove} prompts an LLM without training with the user's spatio-temporal memory, world knowledge about nearby places, and the transitions of other users, GNPR-SID~\cite{gnpr_sid} generates the semantic ID of the next POI, and MELT~\cite{melt} mitigates long-tailed user and item sparsity through mutually enhancing branches.

\begin{figure*}[!ht]
    \centering
    \includegraphics[width=1.0\textwidth,trim={0 0.5bp 0 2bp},clip]{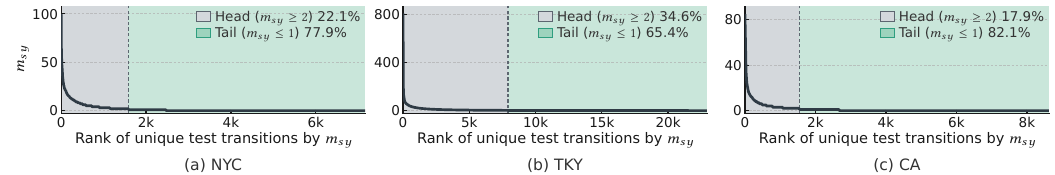}
    \caption{Sorted train-frequency distribution of unique test source--destination transitions. The shaded ranges mark head transitions ($m_{sy}\geq 2$) and tail transitions ($m_{sy}\leq 1$), and the legends report the corresponding percentages.}
    \label{fig:transition_tail_distribution}
\end{figure*}

\begin{figure*}[!ht]
    \centering
    \includegraphics[width=1.0\textwidth,trim={0 0.5bp 0 1.5bp},clip]{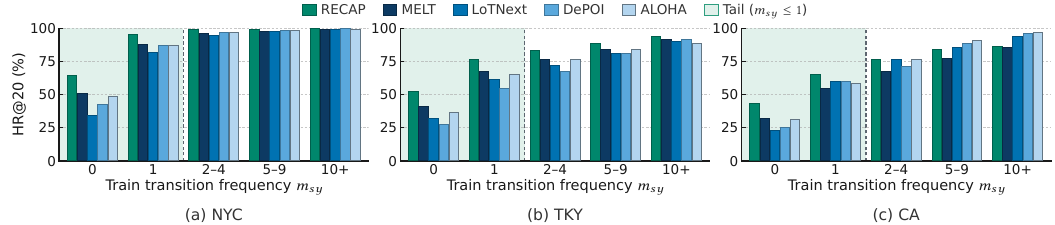}
    \caption{HR@20 of RECAP and the long-tail targeted methods in each bin of the training frequency $m_{s_t y_t}$.}
    \label{fig:tail_frequency_bucket_performance}
\end{figure*}

\subsection{Implementation Details}
\label{app:implementation-details}

We implement RECAP and all baselines with PyTorch and run them on NVIDIA A100 GPUs with 80 GB memory. Following~\cite{flashback,k1poi}, we evaluate all methods under one protocol: every check-in of a test trajectory except the first one is a test sample, and RECAP and all baselines are evaluated on exactly the same test samples, rather than on only the last check-in of each trajectory or one check-in per user. This is closer to a Web service, where the model has to predict the next POI for test samples with both long and short histories. Each baseline keeps the training settings of its original paper, except for the changes described below. For methods that are trained, we report the mean over three random seeds. For every trained method, we keep the checkpoint with the highest validation HR@1 and report its results on the test set.

Hyperparameters are selected on the validation set. For RECAP, the Transformer encoder has 2 layers, hidden dimension 256, 4 attention heads, and feed-forward dimension 512, and the POI and category embeddings have dimensions 128 and 32. We use $k=10$ and $N=2$. In the first stage, we train the backbone for 130 epochs with Adam, a batch size of 512, a learning rate of $3\times10^{-5}$, and a weight decay of $5\times10^{-6}$. In the second stage, the scorer $f$ has two hidden layers of width 64 and is trained for 12 epochs with Adam, a batch size of 256, and a learning rate of $3\times10^{-3}$. Each stage keeps the epoch with the highest validation HR@1. We select $\lambda$ from $\{0.25,0.5,1,2,3,4,6,8,12,16,24,32,48,64,96,128\}$ by the validation HR@1, which gives $\lambda=12$, $8$, and $24$ on NYC, TKY, and CA, respectively.

For the baselines, ROTAN and DPRL use the time of the last input check-in instead of the target visit time, and LLM4POI and AgentMove remove the target visit time from their prompts, so that no method reads the target time. We fix a bug in the released code of GETNext where attention was computed over the batch dimension instead of the sequence dimension. MELT treats POIs as items, and ALOHA is applied to the STHGCN backbone. LLM4POI fine-tunes Llama-2-7B, GNPR-SID generates semantic IDs with Llama-3.1-8B-Instruct, and AgentMove uses Qwen2.5-7B-Instruct.

\section{Additional Experimental Results}
\subsection{Hyperparameters Sensitivity Analysis}
\label{app:hyperparameter-sensitivity}

\subsubsection{Tail Transition Threshold $\eta$.}
\label{app:tail-definition-sensitivity}
The main experiments use $\eta=1$, so tail transitions include the unseen transitions and the transitions that appear once in the training data.
Changing $\eta$ only regroups the test samples, and \textit{does not} determine the model design of RECAP or change its predictions. The gains of RECAP therefore \textit{do not} come from a particular choice of the tail definition.
This threshold follows the dataset distributions in Figure~\ref{fig:transition_tail_distribution}, where the $m_{sy}\leq 1$ region covers 77.9\%, 65.4\%, and 82.1\% of the unique test transitions on NYC, TKY, and CA, respectively.
We also group the test samples by the training frequency $m_{s_t y_t}$ of their transitions into 0, 1, 2--4, 5--9, and 10+, and compare RECAP with the long-tail targeted methods in each bin.
Figure~\ref{fig:tail_frequency_bucket_performance} reports per-dataset HR@20.
RECAP achieves the highest HR@20 in the 0 and 1 bins on all three datasets, and in all bins on NYC and TKY.

On CA, LoTNext, DePOI, and ALOHA are higher in the 5--9 and 10+ bins. The revisit scorer only raises the scores of the POIs that the user has visited (Equation~\eqref{eq:final-logit}). When the target of a frequent transition is a POI that the user has not visited, these visited POIs move ahead of it and can push it out of the top 20, while the baselines do not favor visited POIs. Such samples are most common on CA: they account for 27\%, 11\%, and 1\% of the samples with $m_{s_t y_t}\geq 1$ on CA, TKY, and NYC, respectively. The same adjustment brings the destinations of tail transitions, which are often POIs the user has visited, into the top 20, so $\lambda$ trades off the two effects (Figure~\ref{fig:evidence}(c)).

\begin{table}[!t]
\centering
\caption{Effect of the number of graph hops $N$. Avg. cand. is the average $|C_N(s)|$ over the sources in $\mathcal T_{\mathrm{unseen}}$. H@20 is measured on the test samples whose target the user has not visited before.}
\label{tab:graph_hop_sweep_results}
\small
\setlength{\tabcolsep}{0.55em}
\begin{tabular}{llcccc}
\toprule
Dataset & $N$ & Avg. cand. & Coverage & $10^3\times$SNR & H@20 \\
\midrule
NYC & 1 & 11.5 & 0.00 & 0.000 & 0.0213 \\
 & 2 & 209.0 & 41.26 & \textbf{4.195} & 0.0230 \\
 & 3 & 1737.7 & 77.49 & 0.945 & 0.0254 \\
 & 4 & 4028.0 & 94.90 & 0.499 & 0.0250 \\
 & 5 & 4864.0 & 97.94 & 0.426 & 0.0233 \\
\midrule
TKY & 1 & 21.9 & 0.00 & 0.000 & 0.1392 \\
 & 2 & 1280.9 & 77.12 & \textbf{1.800} & 0.1449 \\
 & 3 & 6215.1 & 97.70 & 0.469 & 0.1450 \\
 & 4 & 7726.8 & 99.71 & 0.385 & 0.1434 \\
 & 5 & 7820.4 & 99.86 & 0.381 & 0.1421 \\
\midrule
CA & 1 & 16.1 & 0.00 & 0.000 & 0.1866 \\
 & 2 & 387.1 & 51.30 & \textbf{2.593} & 0.2131 \\
 & 3 & 2721.8 & 87.47 & 0.628 & 0.2135 \\
 & 4 & 6331.1 & 97.56 & 0.301 & 0.2068 \\
 & 5 & 8922.7 & 99.14 & 0.217 & 0.2014 \\
\bottomrule
\end{tabular}
\end{table}

\subsubsection{Number of Graph Hops $N$.}
\label{app:hops}
For each unseen validation transition $(s,d)\in\mathcal T_{\mathrm{unseen}}$, let $C_N(s)$ be the set of POIs reachable from $s$ within $N$ hops in the transition graph. Let $S_N$ be the number of transitions in $\mathcal T_{\mathrm{unseen}}$ with $d\in C_N(s)$, and let $M_N=\sum_s|C_N(s)|$ be the total number of graph candidates over the sources in $\mathcal T_{\mathrm{unseen}}$. We report the coverage and the signal-to-noise ratio (SNR) of unseen transitions:
\[
\mathrm{Coverage}_N=\frac{S_N}{|\mathcal T_{\mathrm{unseen}}|},
\qquad
\mathrm{SNR}_N=\frac{S_N}{M_N-S_N},
\]
where $M_N-S_N$ counts the candidates that are not unseen validation targets.
At $N=1$, the candidates are only the POIs that $s$ already reaches in training, so the coverage is zero.

Table~\ref{tab:graph_hop_sweep_results} reports the results on the three datasets. From $N=2$ to $N=3$, the average number of graph candidates grows 4.9 to 8.3 times, while the coverage of unseen transitions rises by 21 to 36 percentage points, so most of the added candidates are noise and $\mathrm{SNR}_N$ is the highest at $N=2$ on all three datasets. The last column shows how $N$ affects targets that the user has not visited. We measure it without the revisit scorer, so that it reflects the graph token alone. From $N=1$ to $N=2$, this H@20 rises by 8\%, 4\% and 14\% in relative terms on NYC, TKY and CA. We select $N$ by the SNR on validation data rather than by test accuracy, and RECAP therefore uses $N=2$, which gives the highest SNR.

\begin{table}[!b]
\centering
\caption{HR@1 of methods that use the observed history in different ways. ``+ observed history'' means using the observed history, not only the check-ins in its backbone input.}
\label{tab:history-use}
\small
\begin{tabular}{lccc}
\toprule
Method & NYC & TKY & CA \\
\midrule
K1-POI & 0.2289 & 0.2463 & 0.1596 \\
\quad + observed history & 0.1897 & 0.1192 & 0.1005 \\
\midrule
CaST-POI & 0.2332 & 0.2275 & 0.1524 \\
\quad + observed history in the revisit gate & 0.2534 & 0.2516 & 0.1708 \\
\midrule
RECAP, three statistics of CaST-POI & 0.2392 & 0.2403 & 0.1595 \\
RECAP, six statistics (Table~\ref{tab:revisit-stats}) & \textbf{0.2837} & \textbf{0.2751} & \textbf{0.1963} \\
\bottomrule
\end{tabular}
\end{table}

\begin{figure}[!ht]
    \centering
    \includegraphics[width=\columnwidth,trim={0 2bp 0 2.5bp},clip]{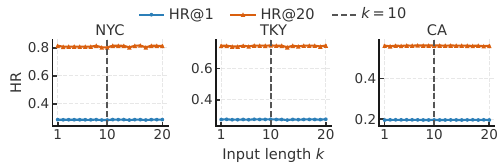}
    \caption{Sensitivity to the number of trajectory tokens encoded by the Transformer backbone. The dashed line marks the $k=10$ setting used in the main experiments.}
    \label{fig:input_window_length}
\end{figure}

\begin{table*}[!ht]
\caption{Efficiency comparison of RECAP and the baselines. Each cell lists the values on NYC / TKY / CA. The best and second-best values on each dataset are in bold and underlined. AgentMove requires no training, and its average rank is computed over the other five columns.}
\label{tab:efficiency}
\small
\setlength{\tabcolsep}{0.35em}
\resizebox{\textwidth}{!}{
\begin{tabular}{lccccccc}
\toprule
Method & H@1 $\uparrow$ & H@10 $\uparrow$ & H@20 $\uparrow$ & \begin{tabular}[c]{@{}c@{}}Total Training\\Time (h) $\downarrow$\end{tabular} & \begin{tabular}[c]{@{}c@{}}Test Evaluation\\Time (s) $\downarrow$\end{tabular} & \begin{tabular}[c]{@{}c@{}}Peak GPU\\Memory (GB) $\downarrow$\end{tabular} & \begin{tabular}[c]{@{}c@{}}Average\\Rank $\downarrow$\end{tabular} \\
\midrule
DeepMove & \phantom{\textbf{0.1526}}\llap{0.1526}\,/\,\phantom{\textbf{0.1809}}\llap{0.1809}\,/\,\phantom{\textbf{0.1026}}\llap{0.1026} & \phantom{\textbf{0.4312}}\llap{0.4312}\,/\,\phantom{\textbf{0.4716}}\llap{0.4716}\,/\,\phantom{\textbf{0.2901}}\llap{0.2901} & \phantom{\textbf{0.4988}}\llap{0.4988}\,/\,\phantom{\textbf{0.5469}}\llap{0.5469}\,/\,\phantom{\textbf{0.3543}}\llap{0.3543} & \phantom{\textbf{13.33}}\llap{0.30}\,/\,\phantom{\textbf{27.79}}\llap{0.89}\,/\,\phantom{\textbf{21.09}}\llap{0.35} & \phantom{\textbf{30650}}\llap{1.63}\,/\,\phantom{\textbf{115121}}\llap{4.25}\,/\,\phantom{\textbf{35890}}\llap{0.93} & \phantom{\textbf{24.59}}\llap{\underline{0.23}}\,/\,\phantom{\textbf{26.60}}\llap{\textbf{0.36}}\,/\,\phantom{\textbf{21.74}}\llap{\underline{0.54}} & \phantom{\textbf{12.00}}\llap{12.00}\,/\,\phantom{\textbf{10.50}}\llap{9.67}\,/\,\phantom{\textbf{10.50}}\llap{10.50} \\
Flashback & \phantom{\textbf{0.1526}}\llap{0.1789}\,/\,\phantom{\textbf{0.1809}}\llap{0.1682}\,/\,\phantom{\textbf{0.1026}}\llap{0.1233} & \phantom{\textbf{0.4312}}\llap{0.5191}\,/\,\phantom{\textbf{0.4716}}\llap{0.4558}\,/\,\phantom{\textbf{0.2901}}\llap{0.3275} & \phantom{\textbf{0.4988}}\llap{0.5924}\,/\,\phantom{\textbf{0.5469}}\llap{0.5273}\,/\,\phantom{\textbf{0.3543}}\llap{0.3889} & \phantom{\textbf{13.33}}\llap{0.34}\,/\,\phantom{\textbf{27.79}}\llap{0.45}\,/\,\phantom{\textbf{21.09}}\llap{0.28} & \phantom{\textbf{30650}}\llap{1.69}\,/\,\phantom{\textbf{115121}}\llap{1.47}\,/\,\phantom{\textbf{35890}}\llap{0.63} & \phantom{\textbf{24.59}}\llap{1.44}\,/\,\phantom{\textbf{26.60}}\llap{2.26}\,/\,\phantom{\textbf{21.74}}\llap{0.57} & \phantom{\textbf{12.00}}\llap{11.00}\,/\,\phantom{\textbf{10.50}}\llap{10.50}\,/\,\phantom{\textbf{10.50}}\llap{8.25} \\
ROTAN & \phantom{\textbf{0.1526}}\llap{0.2185}\,/\,\phantom{\textbf{0.1809}}\llap{0.2467}\,/\,\phantom{\textbf{0.1026}}\llap{0.1482} & \phantom{\textbf{0.4312}}\llap{0.6172}\,/\,\phantom{\textbf{0.4716}}\llap{0.5965}\,/\,\phantom{\textbf{0.2901}}\llap{0.4028} & \phantom{\textbf{0.4988}}\llap{0.6924}\,/\,\phantom{\textbf{0.5469}}\llap{0.6718}\,/\,\phantom{\textbf{0.3543}}\llap{0.4780} & \phantom{\textbf{13.33}}\llap{0.53}\,/\,\phantom{\textbf{27.79}}\llap{1.58}\,/\,\phantom{\textbf{21.09}}\llap{0.79} & \phantom{\textbf{30650}}\llap{0.99}\,/\,\phantom{\textbf{115121}}\llap{2.76}\,/\,\phantom{\textbf{35890}}\llap{0.79} & \phantom{\textbf{24.59}}\llap{24.59}\,/\,\phantom{\textbf{26.60}}\llap{26.60}\,/\,\phantom{\textbf{21.74}}\llap{21.74} & \phantom{\textbf{12.00}}\llap{8.67}\,/\,\phantom{\textbf{10.50}}\llap{6.83}\,/\,\phantom{\textbf{10.50}}\llap{7.17} \\
REPLAY & \phantom{\textbf{0.1526}}\llap{0.1776}\,/\,\phantom{\textbf{0.1809}}\llap{0.1786}\,/\,\phantom{\textbf{0.1026}}\llap{0.1188} & \phantom{\textbf{0.4312}}\llap{0.5123}\,/\,\phantom{\textbf{0.4716}}\llap{0.4715}\,/\,\phantom{\textbf{0.2901}}\llap{0.3212} & \phantom{\textbf{0.4988}}\llap{0.5821}\,/\,\phantom{\textbf{0.5469}}\llap{0.5411}\,/\,\phantom{\textbf{0.3543}}\llap{0.3821} & \phantom{\textbf{13.33}}\llap{0.28}\,/\,\phantom{\textbf{27.79}}\llap{0.40}\,/\,\phantom{\textbf{21.09}}\llap{0.28} & \phantom{\textbf{30650}}\llap{1.56}\,/\,\phantom{\textbf{115121}}\llap{1.31}\,/\,\phantom{\textbf{35890}}\llap{0.65} & \phantom{\textbf{24.59}}\llap{0.80}\,/\,\phantom{\textbf{26.60}}\llap{1.20}\,/\,\phantom{\textbf{21.74}}\llap{0.58} & \phantom{\textbf{12.00}}\llap{11.00}\,/\,\phantom{\textbf{10.50}}\llap{9.00}\,/\,\phantom{\textbf{10.50}}\llap{9.25} \\
DPRL & \phantom{\textbf{0.1526}}\llap{0.1794}\,/\,\phantom{\textbf{0.1809}}\llap{0.2087}\,/\,\phantom{\textbf{0.1026}}\llap{0.1336} & \phantom{\textbf{0.4312}}\llap{0.5159}\,/\,\phantom{\textbf{0.4716}}\llap{0.5417}\,/\,\phantom{\textbf{0.2901}}\llap{0.3789} & \phantom{\textbf{0.4988}}\llap{0.5892}\,/\,\phantom{\textbf{0.5469}}\llap{0.6157}\,/\,\phantom{\textbf{0.3543}}\llap{0.4547} & \phantom{\textbf{13.33}}\llap{\underline{0.08}}\,/\,\phantom{\textbf{27.79}}\llap{\textbf{0.11}}\,/\,\phantom{\textbf{21.09}}\llap{\textbf{0.07}} & \phantom{\textbf{30650}}\llap{0.87}\,/\,\phantom{\textbf{115121}}\llap{0.78}\,/\,\phantom{\textbf{35890}}\llap{0.30} & \phantom{\textbf{24.59}}\llap{0.96}\,/\,\phantom{\textbf{26.60}}\llap{1.36}\,/\,\phantom{\textbf{21.74}}\llap{0.69} & \phantom{\textbf{12.00}}\llap{8.83}\,/\,\phantom{\textbf{10.50}}\llap{5.92}\,/\,\phantom{\textbf{10.50}}\llap{5.50} \\
K1-POI & \phantom{\textbf{0.1526}}\llap{0.2289}\,/\,\phantom{\textbf{0.1809}}\llap{0.2463}\,/\,\phantom{\textbf{0.1026}}\llap{\underline{0.1596}} & \phantom{\textbf{0.4312}}\llap{\underline{0.6364}}\,/\,\phantom{\textbf{0.4716}}\llap{\underline{0.6101}}\,/\,\phantom{\textbf{0.2901}}\llap{\underline{0.4382}} & \phantom{\textbf{0.4988}}\llap{0.7122}\,/\,\phantom{\textbf{0.5469}}\llap{\underline{0.6878}}\,/\,\phantom{\textbf{0.3543}}\llap{\underline{0.5242}} & \phantom{\textbf{13.33}}\llap{\textbf{0.05}}\,/\,\phantom{\textbf{27.79}}\llap{\underline{0.15}}\,/\,\phantom{\textbf{21.09}}\llap{\underline{0.09}} & \phantom{\textbf{30650}}\llap{\textbf{0.05}}\,/\,\phantom{\textbf{115121}}\llap{\textbf{0.22}}\,/\,\phantom{\textbf{35890}}\llap{\textbf{0.08}} & \phantom{\textbf{24.59}}\llap{2.72}\,/\,\phantom{\textbf{26.60}}\llap{2.79}\,/\,\phantom{\textbf{21.74}}\llap{2.83} & \phantom{\textbf{12.00}}\llap{\underline{3.67}}\,/\,\phantom{\textbf{10.50}}\llap{\underline{3.33}}\,/\,\phantom{\textbf{10.50}}\llap{\underline{3.00}} \\
\midrule
GETNext & \phantom{\textbf{0.1526}}\llap{0.1766}\,/\,\phantom{\textbf{0.1809}}\llap{0.1503}\,/\,\phantom{\textbf{0.1026}}\llap{0.1192} & \phantom{\textbf{0.4312}}\llap{0.4690}\,/\,\phantom{\textbf{0.4716}}\llap{0.3350}\,/\,\phantom{\textbf{0.2901}}\llap{0.2849} & \phantom{\textbf{0.4988}}\llap{0.5413}\,/\,\phantom{\textbf{0.5469}}\llap{0.3832}\,/\,\phantom{\textbf{0.3543}}\llap{0.3134} & \phantom{\textbf{13.33}}\llap{6.70}\,/\,\phantom{\textbf{27.79}}\llap{27.79}\,/\,\phantom{\textbf{21.09}}\llap{21.09} & \phantom{\textbf{30650}}\llap{4.13}\,/\,\phantom{\textbf{115121}}\llap{13.84}\,/\,\phantom{\textbf{35890}}\llap{4.10} & \phantom{\textbf{24.59}}\llap{2.25}\,/\,\phantom{\textbf{26.60}}\llap{3.36}\,/\,\phantom{\textbf{21.74}}\llap{5.85} & \phantom{\textbf{12.00}}\llap{14.67}\,/\,\phantom{\textbf{10.50}}\llap{14.67}\,/\,\phantom{\textbf{10.50}}\llap{14.00} \\
STHGCN & \phantom{\textbf{0.1526}}\llap{0.2290}\,/\,\phantom{\textbf{0.1809}}\llap{\underline{0.2485}}\,/\,\phantom{\textbf{0.1026}}\llap{0.1461} & \phantom{\textbf{0.4312}}\llap{0.6255}\,/\,\phantom{\textbf{0.4716}}\llap{0.5915}\,/\,\phantom{\textbf{0.2901}}\llap{0.4064} & \phantom{\textbf{0.4988}}\llap{0.6968}\,/\,\phantom{\textbf{0.5469}}\llap{0.6660}\,/\,\phantom{\textbf{0.3543}}\llap{0.4864} & \phantom{\textbf{13.33}}\llap{0.73}\,/\,\phantom{\textbf{27.79}}\llap{8.18}\,/\,\phantom{\textbf{21.09}}\llap{2.99} & \phantom{\textbf{30650}}\llap{3.15}\,/\,\phantom{\textbf{115121}}\llap{46.98}\,/\,\phantom{\textbf{35890}}\llap{6.35} & \phantom{\textbf{24.59}}\llap{4.43}\,/\,\phantom{\textbf{26.60}}\llap{13.76}\,/\,\phantom{\textbf{21.74}}\llap{9.02} & \phantom{\textbf{12.00}}\llap{8.25}\,/\,\phantom{\textbf{10.50}}\llap{8.50}\,/\,\phantom{\textbf{10.50}}\llap{8.25} \\
DCHL & \phantom{\textbf{0.1526}}\llap{0.1841}\,/\,\phantom{\textbf{0.1809}}\llap{0.1301}\,/\,\phantom{\textbf{0.1026}}\llap{0.1025} & \phantom{\textbf{0.4312}}\llap{0.6199}\,/\,\phantom{\textbf{0.4716}}\llap{0.4970}\,/\,\phantom{\textbf{0.2901}}\llap{0.3745} & \phantom{\textbf{0.4988}}\llap{\underline{0.7216}}\,/\,\phantom{\textbf{0.5469}}\llap{0.5919}\,/\,\phantom{\textbf{0.3543}}\llap{0.4616} & \phantom{\textbf{13.33}}\llap{0.39}\,/\,\phantom{\textbf{27.79}}\llap{1.98}\,/\,\phantom{\textbf{21.09}}\llap{2.90} & \phantom{\textbf{30650}}\llap{1.08}\,/\,\phantom{\textbf{115121}}\llap{6.21}\,/\,\phantom{\textbf{35890}}\llap{2.63} & \phantom{\textbf{24.59}}\llap{0.71}\,/\,\phantom{\textbf{26.60}}\llap{1.67}\,/\,\phantom{\textbf{21.74}}\llap{2.59} & \phantom{\textbf{12.00}}\llap{6.50}\,/\,\phantom{\textbf{10.50}}\llap{11.33}\,/\,\phantom{\textbf{10.50}}\llap{9.92} \\
MSAHG & \phantom{\textbf{0.1526}}\llap{0.1851}\,/\,\phantom{\textbf{0.1809}}\llap{0.1400}\,/\,\phantom{\textbf{0.1026}}\llap{0.1118} & \phantom{\textbf{0.4312}}\llap{0.6036}\,/\,\phantom{\textbf{0.4716}}\llap{0.4964}\,/\,\phantom{\textbf{0.2901}}\llap{0.3542} & \phantom{\textbf{0.4988}}\llap{0.7071}\,/\,\phantom{\textbf{0.5469}}\llap{0.5963}\,/\,\phantom{\textbf{0.3543}}\llap{0.4228} & \phantom{\textbf{13.33}}\llap{0.52}\,/\,\phantom{\textbf{27.79}}\llap{0.51}\,/\,\phantom{\textbf{21.09}}\llap{5.95} & \phantom{\textbf{30650}}\llap{0.90}\,/\,\phantom{\textbf{115121}}\llap{4.12}\,/\,\phantom{\textbf{35890}}\llap{1.60} & \phantom{\textbf{24.59}}\llap{20.85}\,/\,\phantom{\textbf{26.60}}\llap{31.41}\,/\,\phantom{\textbf{21.74}}\llap{39.78} & \phantom{\textbf{12.00}}\llap{8.67}\,/\,\phantom{\textbf{10.50}}\llap{11.50}\,/\,\phantom{\textbf{10.50}}\llap{12.50} \\
\midrule
LLM4POI & \phantom{\textbf{0.1526}}\llap{0.2077}\,/\,\phantom{\textbf{0.1809}}\llap{0.1954}\,/\,\phantom{\textbf{0.1026}}\llap{0.1253} & \phantom{\textbf{0.4312}}\llap{0.5995}\,/\,\phantom{\textbf{0.4716}}\llap{0.5255}\,/\,\phantom{\textbf{0.2901}}\llap{0.3415} & \phantom{\textbf{0.4988}}\llap{0.6461}\,/\,\phantom{\textbf{0.5469}}\llap{0.5757}\,/\,\phantom{\textbf{0.3543}}\llap{0.3722} & \phantom{\textbf{13.33}}\llap{13.33}\,/\,\phantom{\textbf{27.79}}\llap{41.23}\,/\,\phantom{\textbf{21.09}}\llap{18.78} & \phantom{\textbf{30650}}\llap{5726}\,/\,\phantom{\textbf{115121}}\llap{20063}\,/\,\phantom{\textbf{35890}}\llap{6240} & \phantom{\textbf{24.59}}\llap{31.30}\,/\,\phantom{\textbf{26.60}}\llap{31.50}\,/\,\phantom{\textbf{21.74}}\llap{41.91} & \phantom{\textbf{12.00}}\llap{13.33}\,/\,\phantom{\textbf{10.50}}\llap{13.67}\,/\,\phantom{\textbf{10.50}}\llap{14.33} \\
AgentMove & \phantom{\textbf{0.1526}}\llap{0.1626}\,/\,\phantom{\textbf{0.1809}}\llap{0.1266}\,/\,\phantom{\textbf{0.1026}}\llap{0.0907} & \phantom{\textbf{0.4312}}\llap{0.5499}\,/\,\phantom{\textbf{0.4716}}\llap{0.4486}\,/\,\phantom{\textbf{0.2901}}\llap{0.2865} & \phantom{\textbf{0.4988}}\llap{0.5828}\,/\,\phantom{\textbf{0.5469}}\llap{0.4895}\,/\,\phantom{\textbf{0.3543}}\llap{0.3077} & \ooalign{\phantom{\textbf{13.33}}\cr\hfil--\hfil\cr}\,/\,\ooalign{\phantom{\textbf{27.79}}\cr\hfil--\hfil\cr}\,/\,\ooalign{\phantom{\textbf{21.09}}\cr\hfil--\hfil\cr} & \phantom{\textbf{30650}}\llap{30650}\,/\,\phantom{\textbf{115121}}\llap{115121}\,/\,\phantom{\textbf{35890}}\llap{35890} & \phantom{\textbf{24.59}}\llap{25.29}\,/\,\phantom{\textbf{26.60}}\llap{27.22}\,/\,\phantom{\textbf{21.74}}\llap{26.33} & \phantom{\textbf{12.00}}\llap{15.60}\,/\,\phantom{\textbf{10.50}}\llap{17.00}\,/\,\phantom{\textbf{10.50}}\llap{17.00} \\
GNPR-SID & \phantom{\textbf{0.1526}}\llap{0.2230}\,/\,\phantom{\textbf{0.1809}}\llap{0.2108}\,/\,\phantom{\textbf{0.1026}}\llap{0.1548} & \phantom{\textbf{0.4312}}\llap{0.6134}\,/\,\phantom{\textbf{0.4716}}\llap{0.5240}\,/\,\phantom{\textbf{0.2901}}\llap{0.3725} & \phantom{\textbf{0.4988}}\llap{0.6558}\,/\,\phantom{\textbf{0.5469}}\llap{0.5897}\,/\,\phantom{\textbf{0.3543}}\llap{0.4200} & \phantom{\textbf{13.33}}\llap{12.96}\,/\,\phantom{\textbf{27.79}}\llap{49.91}\,/\,\phantom{\textbf{21.09}}\llap{27.34} & \phantom{\textbf{30650}}\llap{1147}\,/\,\phantom{\textbf{115121}}\llap{3958}\,/\,\phantom{\textbf{35890}}\llap{1196} & \phantom{\textbf{24.59}}\llap{52.84}\,/\,\phantom{\textbf{26.60}}\llap{53.03}\,/\,\phantom{\textbf{21.74}}\llap{53.25} & \phantom{\textbf{12.00}}\llap{12.17}\,/\,\phantom{\textbf{10.50}}\llap{13.50}\,/\,\phantom{\textbf{10.50}}\llap{12.50} \\
\midrule
MELT & \phantom{\textbf{0.1526}}\llap{0.2121}\,/\,\phantom{\textbf{0.1809}}\llap{0.2253}\,/\,\phantom{\textbf{0.1026}}\llap{0.1280} & \phantom{\textbf{0.4312}}\llap{0.6308}\,/\,\phantom{\textbf{0.4716}}\llap{0.5979}\,/\,\phantom{\textbf{0.2901}}\llap{0.3818} & \phantom{\textbf{0.4988}}\llap{0.7148}\,/\,\phantom{\textbf{0.5469}}\llap{0.6771}\,/\,\phantom{\textbf{0.3543}}\llap{0.4616} & \phantom{\textbf{13.33}}\llap{1.19}\,/\,\phantom{\textbf{27.79}}\llap{3.91}\,/\,\phantom{\textbf{21.09}}\llap{2.29} & \phantom{\textbf{30650}}\llap{0.22}\,/\,\phantom{\textbf{115121}}\llap{0.78}\,/\,\phantom{\textbf{35890}}\llap{0.24} & \phantom{\textbf{24.59}}\llap{1.11}\,/\,\phantom{\textbf{26.60}}\llap{1.23}\,/\,\phantom{\textbf{21.74}}\llap{1.02} & \phantom{\textbf{12.00}}\llap{6.00}\,/\,\phantom{\textbf{10.50}}\llap{5.25}\,/\,\phantom{\textbf{10.50}}\llap{6.58} \\
LoTNext & \phantom{\textbf{0.1526}}\llap{0.1956}\,/\,\phantom{\textbf{0.1809}}\llap{0.2141}\,/\,\phantom{\textbf{0.1026}}\llap{0.1380} & \phantom{\textbf{0.4312}}\llap{0.5403}\,/\,\phantom{\textbf{0.4716}}\llap{0.5541}\,/\,\phantom{\textbf{0.2901}}\llap{0.3662} & \phantom{\textbf{0.4988}}\llap{0.6144}\,/\,\phantom{\textbf{0.5469}}\llap{0.6252}\,/\,\phantom{\textbf{0.3543}}\llap{0.4368} & \phantom{\textbf{13.33}}\llap{0.27}\,/\,\phantom{\textbf{27.79}}\llap{0.67}\,/\,\phantom{\textbf{21.09}}\llap{0.50} & \phantom{\textbf{30650}}\llap{1.06}\,/\,\phantom{\textbf{115121}}\llap{2.94}\,/\,\phantom{\textbf{35890}}\llap{1.54} & \phantom{\textbf{24.59}}\llap{\textbf{0.13}}\,/\,\phantom{\textbf{26.60}}\llap{\underline{0.38}}\,/\,\phantom{\textbf{21.74}}\llap{\textbf{0.35}} & \phantom{\textbf{12.00}}\llap{7.67}\,/\,\phantom{\textbf{10.50}}\llap{6.17}\,/\,\phantom{\textbf{10.50}}\llap{7.33} \\
DePOI & \phantom{\textbf{0.1526}}\llap{0.1783}\,/\,\phantom{\textbf{0.1809}}\llap{0.1518}\,/\,\phantom{\textbf{0.1026}}\llap{0.0828} & \phantom{\textbf{0.4312}}\llap{0.5950}\,/\,\phantom{\textbf{0.4716}}\llap{0.5112}\,/\,\phantom{\textbf{0.2901}}\llap{0.3511} & \phantom{\textbf{0.4988}}\llap{0.6716}\,/\,\phantom{\textbf{0.5469}}\llap{0.5983}\,/\,\phantom{\textbf{0.3543}}\llap{0.4469} & \phantom{\textbf{13.33}}\llap{3.73}\,/\,\phantom{\textbf{27.79}}\llap{21.28}\,/\,\phantom{\textbf{21.09}}\llap{22.57} & \phantom{\textbf{30650}}\llap{6.23}\,/\,\phantom{\textbf{115121}}\llap{28.15}\,/\,\phantom{\textbf{35890}}\llap{11.21} & \phantom{\textbf{24.59}}\llap{4.93}\,/\,\phantom{\textbf{26.60}}\llap{7.68}\,/\,\phantom{\textbf{21.74}}\llap{10.19} & \phantom{\textbf{12.00}}\llap{12.67}\,/\,\phantom{\textbf{10.50}}\llap{12.00}\,/\,\phantom{\textbf{10.50}}\llap{13.83} \\
ALOHA & \phantom{\textbf{0.1526}}\llap{\underline{0.2298}}\,/\,\phantom{\textbf{0.1809}}\llap{0.2258}\,/\,\phantom{\textbf{0.1026}}\llap{0.1423} & \phantom{\textbf{0.4312}}\llap{0.6241}\,/\,\phantom{\textbf{0.4716}}\llap{0.5654}\,/\,\phantom{\textbf{0.2901}}\llap{0.4067} & \phantom{\textbf{0.4988}}\llap{0.7003}\,/\,\phantom{\textbf{0.5469}}\llap{0.6480}\,/\,\phantom{\textbf{0.3543}}\llap{0.4878} & \phantom{\textbf{13.33}}\llap{0.75}\,/\,\phantom{\textbf{27.79}}\llap{8.23}\,/\,\phantom{\textbf{21.09}}\llap{3.02} & \phantom{\textbf{30650}}\llap{3.12}\,/\,\phantom{\textbf{115121}}\llap{46.37}\,/\,\phantom{\textbf{35890}}\llap{6.34} & \phantom{\textbf{24.59}}\llap{4.43}\,/\,\phantom{\textbf{26.60}}\llap{13.77}\,/\,\phantom{\textbf{21.74}}\llap{9.02} & \phantom{\textbf{12.00}}\llap{8.08}\,/\,\phantom{\textbf{10.50}}\llap{9.50}\,/\,\phantom{\textbf{10.50}}\llap{8.08} \\
\midrule
RECAP & \phantom{\textbf{0.1526}}\llap{\textbf{0.2837}}\,/\,\phantom{\textbf{0.1809}}\llap{\textbf{0.2751}}\,/\,\phantom{\textbf{0.1026}}\llap{\textbf{0.1963}} & \phantom{\textbf{0.4312}}\llap{\textbf{0.7402}}\,/\,\phantom{\textbf{0.4716}}\llap{\textbf{0.6727}}\,/\,\phantom{\textbf{0.2901}}\llap{\textbf{0.4829}} & \phantom{\textbf{0.4988}}\llap{\textbf{0.8025}}\,/\,\phantom{\textbf{0.5469}}\llap{\textbf{0.7485}}\,/\,\phantom{\textbf{0.3543}}\llap{\textbf{0.5615}} & \phantom{\textbf{13.33}}\llap{0.19}\,/\,\phantom{\textbf{27.79}}\llap{0.74}\,/\,\phantom{\textbf{21.09}}\llap{0.34} & \phantom{\textbf{30650}}\llap{\underline{0.16}}\,/\,\phantom{\textbf{115121}}\llap{\underline{0.62}}\,/\,\phantom{\textbf{35890}}\llap{\underline{0.22}} & \phantom{\textbf{24.59}}\llap{0.53}\,/\,\phantom{\textbf{26.60}}\llap{0.74}\,/\,\phantom{\textbf{21.74}}\llap{1.10} & \phantom{\textbf{12.00}}\llap{\textbf{1.83}}\,/\,\phantom{\textbf{10.50}}\llap{\textbf{2.50}}\,/\,\phantom{\textbf{10.50}}\llap{\textbf{2.83}} \\
\bottomrule
\end{tabular}
}
\end{table*}

\subsubsection{Input Window Length $k$.}
\label{app:input-window-length}
The backbone reads the most recent $k$ check-ins of the observed history (Section~\ref{sec:method}). We vary $k$ from $1$ to $20$ and evaluate every value of $k$ on the same test samples.
Figure~\ref{fig:input_window_length} shows that the HR@1 and HR@20 of RECAP change little across $k$ on all three datasets.
The revisit scorer reads statistics of the user's history up to the current check-in, not only the check-ins in the input window, so a short window loses little evidence.
This agrees with K1-POI~\cite{k1poi}, the strongest baseline, whose sequence model reads only the last check-in. Earlier check-ins are therefore better used through the statistics of the revisit scorer than through a longer input window, as Table~\ref{tab:history-use} further shows.
We use $k=10$ in the main experiments.

Table~\ref{tab:history-use} examines how the observed history $\mathcal H_t$ (Section~\ref{sec:method}) should be used. First, encoding it only is not enough: K1-POI, the strongest baseline, becomes less accurate on all three datasets when it encodes the observed history instead of only the last check-in. STHGCN and ALOHA, which link each prediction to the user's earlier trajectories in the training data (Section~\ref{sec:evidence-analysis}), also stay below RECAP. Second, a method that scores visited POIs explicitly needs the observed history: CaST-POI~\cite{castpoi}, a concurrent work, scores visited POIs with a revisit gate, and its HR@1 rises on all three datasets when the gate reads the observed history rather than only the check-ins in its default input, although it remains below RECAP. Third, reading the observed history is not enough either if visited POIs are not scored under the current context: within RECAP, we keep the backbone, the check-ins read by the revisit scorer, and its training fixed, and only replace the six statistics in Table~\ref{tab:revisit-stats} with the three used by CaST-POI, namely the number of visits, how recently the POI was visited, and whether it has been visited, which lowers HR@1 by 0.035 to 0.045.

In summary, using the observed history effectively requires two things: the model has to read the observed history, and since tail transitions give the parameters little training signal, it cannot leave the history to be learned implicitly but has to make better use of it.

\subsection{Error Analysis}
\label{app:error-analysis}

We analyze the errors of RECAP on the 8,775 NYC test samples whose target is a POI that the user has visited. When RECAP misses the target, its top-1 POI is always another POI that the user has visited, so its errors come from choosing among the visited POIs. People mostly return to POIs that they visit often or have visited recently~\cite{song2010,barbosa2015}, and the errors concentrate on targets that the user rarely visits. We call a target rarely visited if at most 5\% of the user's earlier check-ins are at it (Table~\ref{tab:error-analysis}). Rarely visited targets make up 33\% of the samples but 45\% of the errors of RECAP. These targets are equally hard for the other methods, and RECAP still achieves a higher HR@1 than ALOHA, the strongest baseline on these samples.

\begin{table}[!ht]
\centering
\caption{Errors of RECAP on the NYC test samples whose target is a POI that the user has visited. A target is rarely visited if at most 5\% of the user's earlier check-ins are at it. Errors is the share of the HR@1 errors of RECAP.}
\label{tab:error-analysis}
\small
\begin{tabular}{lrrc}
\toprule
Target & Samples & Errors & HR@1 (RECAP / ALOHA) \\
\midrule
Rarely visited & 33.3\% & 45.4\% & 0.091 / 0.035 \\
Others & 66.7\% & 54.6\% & 0.454 / 0.387 \\
\bottomrule
\end{tabular}
\end{table}

\subsection{Additional Efficiency Analysis}
\label{app:efficiency-analysis}

Table~\ref{tab:efficiency} compares the efficiency of RECAP and the baselines. HR@1, HR@10, and HR@20 are the three-seed means reported in Table~\ref{tab:main_results}, total training time and peak PyTorch-allocated GPU memory are averaged over the same three seeds on a single A100 GPU, and test evaluation time measures one full test-set evaluation at the best validation checkpoint of one fixed seed. For the three LLM-based baselines, peak GPU memory is measured in a separate run with the same configuration on the same GPU. Average rank is computed over these six columns.
RECAP obtains the best average rank on all three datasets. It achieves the best HR@1, HR@10, and HR@20, while its training takes less than one hour and its peak GPU memory stays below 1.2 GB on all three datasets.
The three LLM-based baselines are strong on sparse transitions: LLM4POI has the highest HR@1 among all baselines on both tail and unseen transitions of all three datasets (Table~\ref{tab:full_head_tail_results}). However, they take about 0.1 to 3.4 seconds per test sample and 25 to 53 GB of GPU memory, which makes them difficult to deploy in online Web recommendation services, whereas RECAP evaluates an entire test set in less than one second.

\section{Limitations and Future Work}
\label{app:limitations-broader-impacts}

This paper studies transition-level long-tail modeling under the data processing of prior work. Our processing follows \citet{getnext} and keeps in the validation and test splits only the users and POIs that also appear in the training split, and each model is trained and tested within one city. Our experiments therefore do not cover users without any history, POIs that never appear in the training data, or transfer to cities that are not seen in training. These settings make tail and unseen transitions more severe, and extending transition-level modeling to them is an important direction for future work. Moreover, because the revisit bonus ranks visited POIs ahead of the others, RECAP ranks targets that the user has not visited lower than targets that the user has visited before. Estimating for each case whether the user will revisit is left for future work.

\end{document}